\documentclass[pre,twocolumn,showpacs]{revtex4}
\usepackage{graphicx}
\usepackage{amsmath}
\usepackage{amssymb}
\usepackage{bm}
\usepackage{latexsym}

\begin{document}

\title{The Growing Correlation Length in Glasses}
\author{C. J. Fullerton}
\altaffiliation{current address: Department of Physics, University of Bath, Bath, BA2 7AY, UK}
\affiliation{School of Physics and Astronomy, University of Manchester,
Manchester M13 9PL, UK}
\author{M. A. Moore}
\affiliation{School of Physics and Astronomy, University of Manchester,
Manchester M13 9PL, UK}

\date{\today}

\begin{abstract}
 The  growing correlation  length observed  in supercooled  liquids as
 their  temperature is  lowered has  been studied  with the  aid  of a
 single occupancy cell model. This  model becomes more accurate as the
 density of the system is increased.  One of its advantages is that it
 permits  a simple mapping  to a  spin system  and the  effective spin
 Hamiltonian is  easily obtained for  smooth interparticle potentials.
 For a binary liquid mixture  the effective spin Hamiltonian is in the
 universality  class of the  Ising spin  glass in  a field.   No phase
 transition  at  finite temperatures  is  therefore  expected and  the
 correlation   length   will   stay   finite  right   down   to   zero
 temperature. For  binary mixtures of  hard disks and spheres  we were
 not able  to obtain the effective spin  Hamiltonian analytically, but
 have  done  simulations  to obtain  its  form.  It  again is  in  the
 universality class  of the Ising spin  glass in a  field. However, in
 this  case the effective  field can  be shown  to go  to zero  at the
 density of maximum  packing in the model, (which is  close to that of
 random close  packing), which means that the  correlation length will
 diverge as  the density approaches  its maximum.  The  exponent $\nu$
 describing the divergence  is related in $d$ dimensions  to the Ising
 spin  glass  domain wall  energy  exponent  $\theta$ via  $\nu=2/(d-2
 \theta)$.
\end{abstract}
\maketitle

\section{Introduction}

One of the key concepts which has emerged in the last few years in the
field of glasses  is that of a growing  correlation length scale $\xi$
\cite{Berthier05,  Cavagna, Biroli,Kob,  Berthier:2011tk}.   There are
now  many  ways  of  defining  and  obtaining  such  a  length  scale:
point-to-set   \cite{Cavagna},    patches   \cite{Kurchan},   dynamics
\cite{Berthier05, Kob} etc.  When it becomes large, they are probably
all proportional to  each other, as they are  basically just a measure
of the  size of the  cooperatively re-arranging regions in  the liquid
\cite{Kob}.  Simulations show that  $\xi$ increases as the temperature
decreases, or  in the case  of hard sphere  and hard disk  systems, as
their density is increased. In this paper we report on our attempts to
understand  this growth,  particularly  in the  context  of hard  disk
systems  in two  dimensions but  also for  particles  interacting with
realistic potentials in any dimension.

The leading theory  for the growth of the  correlation length has been
that of the Random First-Order Transition (RFOT) theory \cite{KTW, LW,
  BB}.   In  this  theory  the  growth is  driven  by  the  decreasing
configurational entropy or complexity \cite{MP, PZ} of the supercooled
liquid  as its temperature  is decreased  towards $T_K$,  the Kauzmann
temperature  \cite{Kauzmann}.   In hard  spheres  there  is a  packing
fraction $\phi_K$ at which the  complexity apparently goes to zero, at
least  in  the mean-field  calculations  of Refs.   \cite{MP,
  PZ}. At  this density the  correlation length diverges  to infinity.
However, there  are arguments that  RFOT theory must be  incorrect for
systems in any finite dimension \cite{YM12b}.

\begin{figure}
\begin{center}
	\resizebox{67mm}{!}{\includegraphics{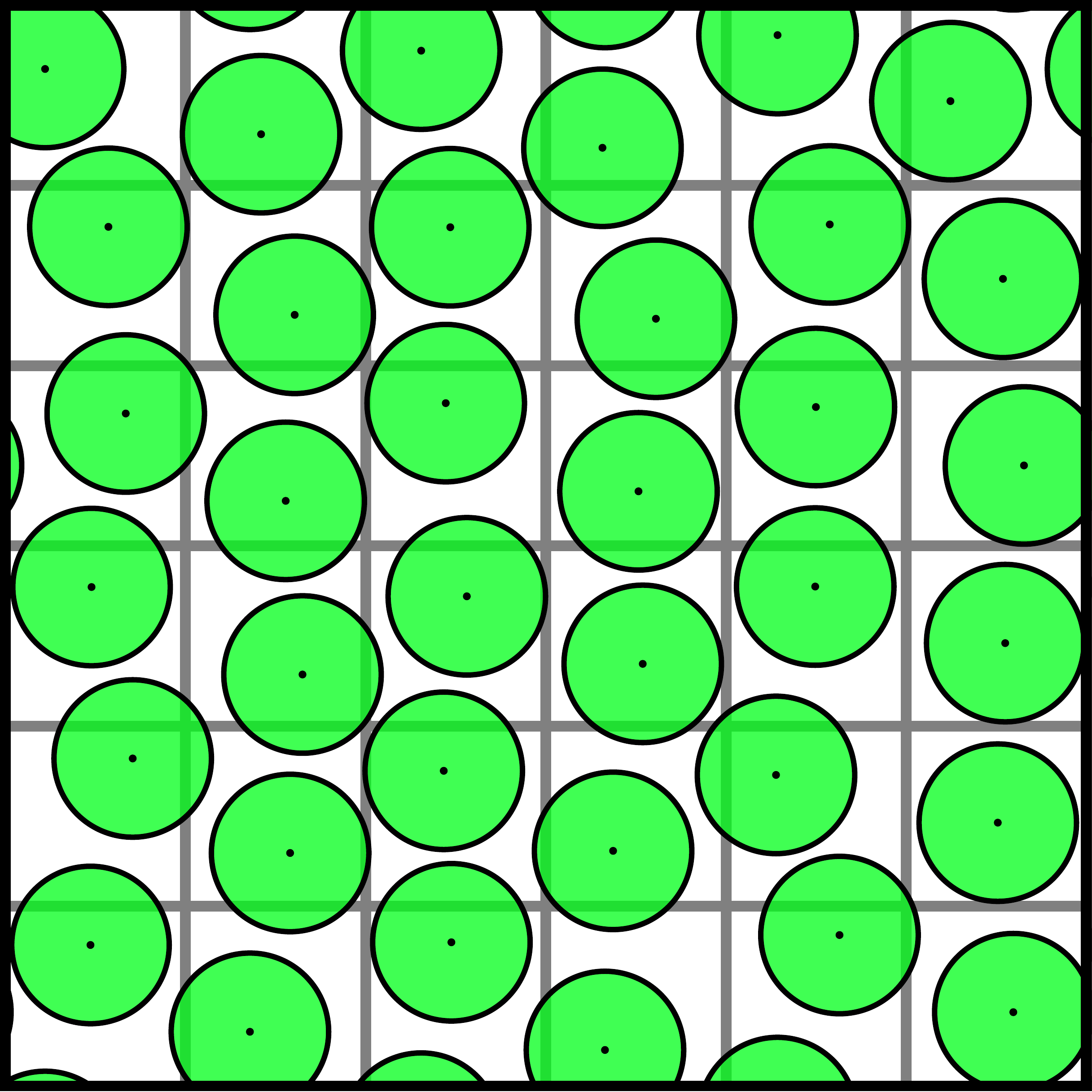}}
	\caption[A  hard  disk   system  with  single  cell  occupancy
          constratins]{  (Color  online)  The  hard disk  system  with
          single  cell occupancy  constraints. The  square  cells have
          grey outlines, and each cell  can contain the center of only
          one  disk (these  are marked  as black  points).   The outer
          edges of the disks do  not interact with the cell walls, but
          only with the outer  edges of other disks. Periodic boundary
          conditions have been used here and throughout this paper.}
\label{MonaSOCM}
\end{center}
\end{figure}

In this  paper we  shall try  to understand the  growth of  the length
scale $\xi$  not on the  basis of RFOT  theory but from  lessons which
have been learnt from studying  in finite dimensions the same $p$-spin
models which  inspired the RFOT theory.  In  Refs. \cite{YM12b,MD, YM}
it has been shown that  these models behave at low temperatures rather
like  an Ising  spin  glass  in a  field  \cite{TM}.  Furthermore  the
correlation length grows as the temperature is decreased but saturates
to  a finite  value at  $T =  0$. It  has also  been argued  that real
glasses as well as $p$-spin models behave like Ising spin glasses in a
field \cite{Yeo}.  This approach involved extensive use of the replica
trick and  is quite non-intuitive. It  is one of the  purposes of this
paper to  explain why, say, a  binary mixture of hard  spheres at high
densities will have features in  common with Ising spin glasses in the
presence of  a field, but  without the aid  of the heavy  machinery of
replicas.

\begin{figure}
\begin{center}
	\resizebox{67mm}{!}{\includegraphics{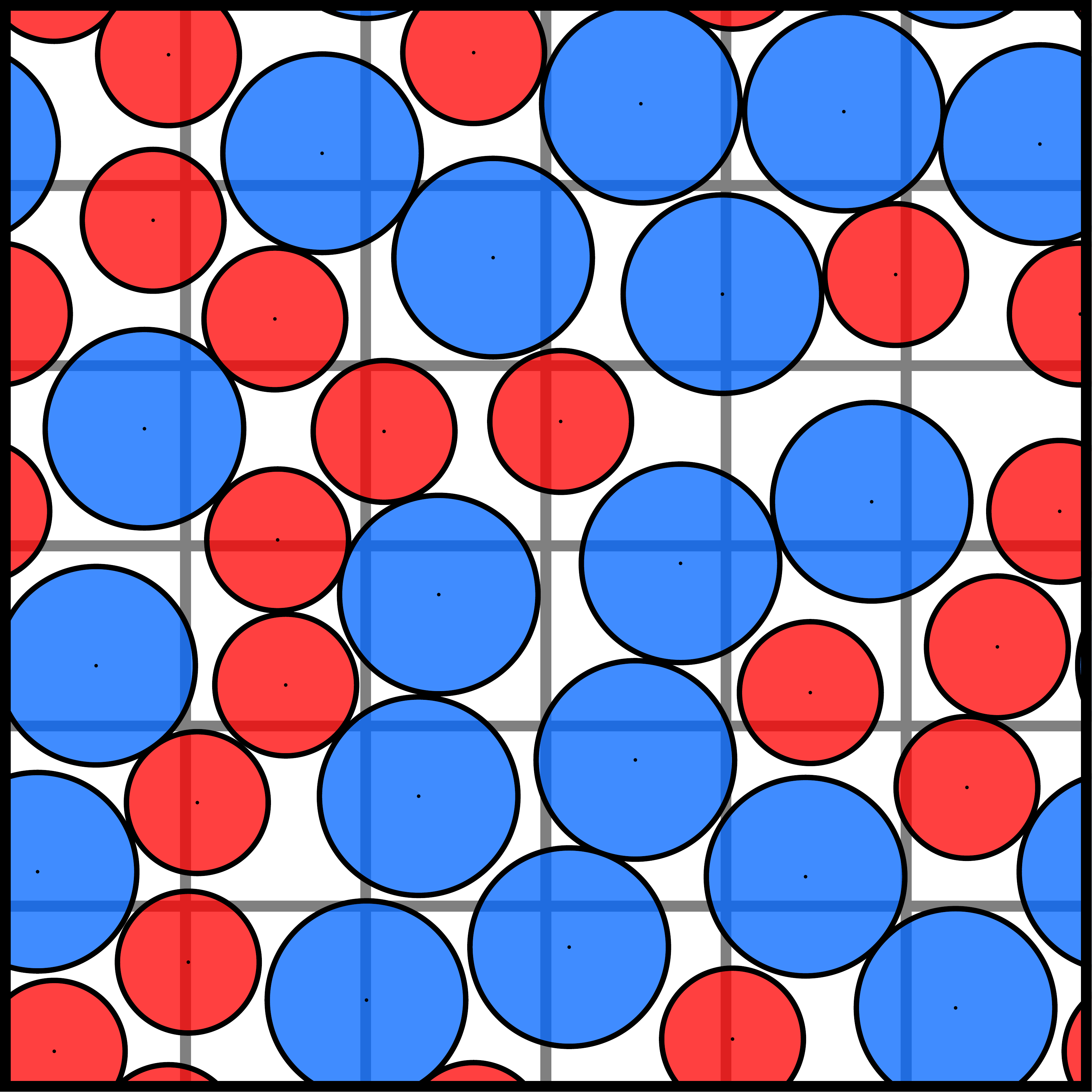}} \\
	 \resizebox{67mm}{!}{\includegraphics{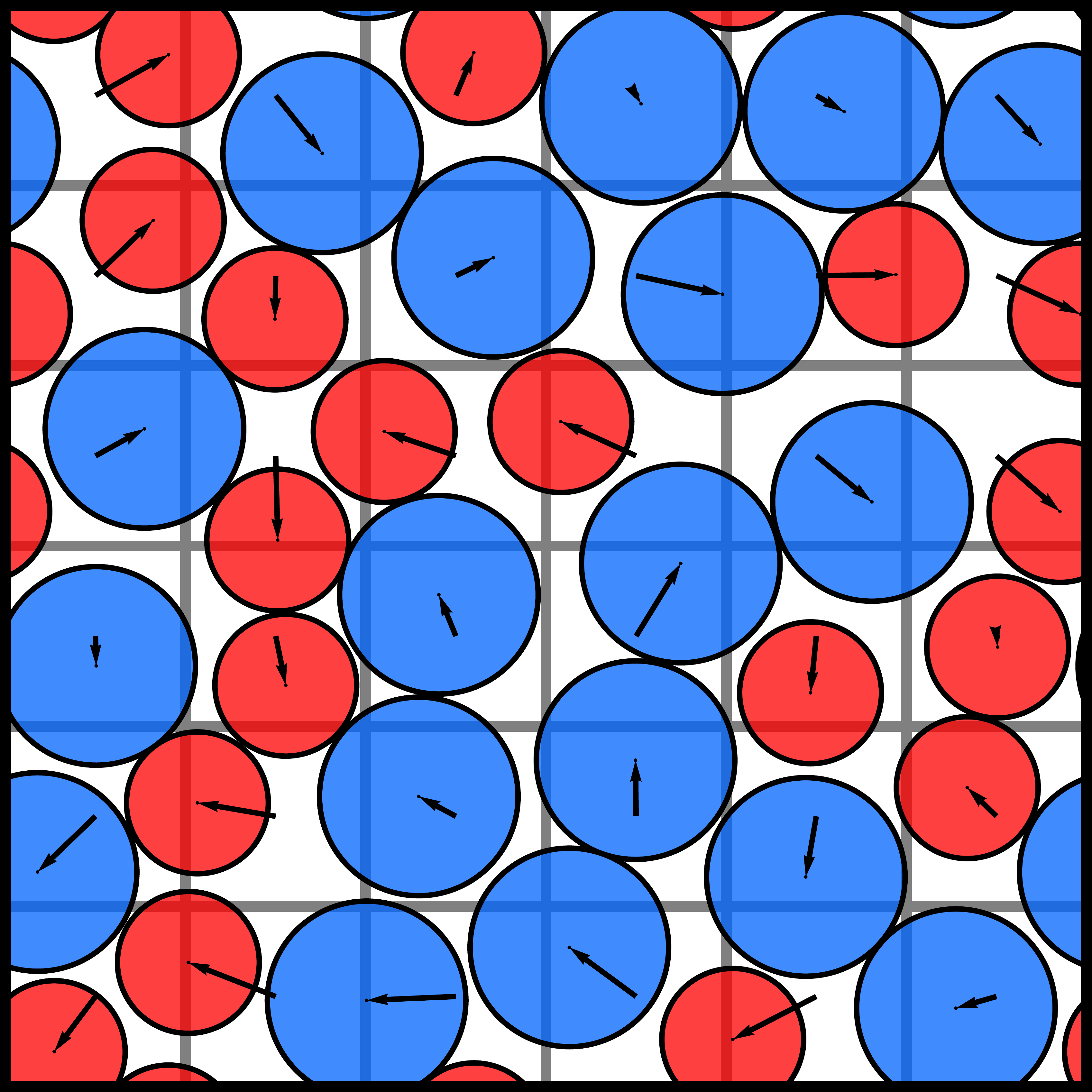}} \\
         \resizebox{67mm}{!}{\includegraphics{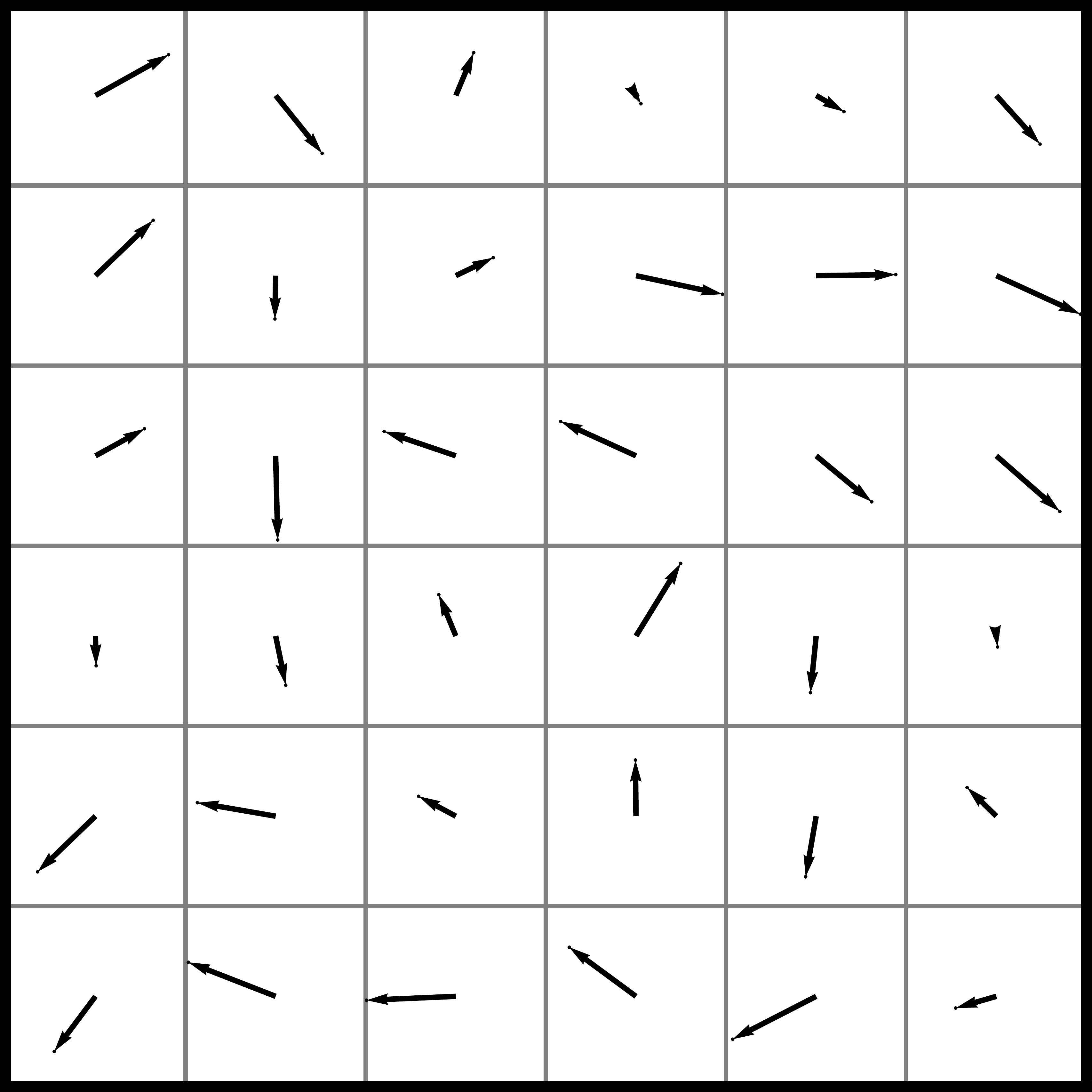}}
	\caption[ The  spin  mapping  in  action]{ (Color online)
 The spin  mapping  in
          action. Vectors  are drawn from  the center of each  cell to
          the center of  the disk occupying that cell  - these vectors
          are  the  spins.  The  disks  are then  forgotten,  and  the
           system  is treated as a  spin
          system. Note that  while there are two types  of disk (large
          and  small)  there  is  only  one type  of  spin  after  the
          mapping.  The  spin  system  contains quenched  disorder,  a
          consequence  of the fact  that each  disk is  constrained to
          remain in the cell to which it is first assigned.} \label{spinmapping}
\end{center}
\end{figure}

To this end, we introduce in Sec. \ref{SOCM} the Single Occupancy Cell
(SOC) model \cite{Hoover:1966cf,  Hoover:1968ux}. In two dimensions it
is a  model in which the  \textit{centers} of the hard  disks are each
\textit{constrained} to stay \textit{forever}  within a plaquette of a
square lattice grid as  in Fig. \ref{MonaSOCM}. (The generalization of
this to  higher dimensions is simple:  in $d=3$ one  would use spheres
whose centers  are confined  within the primitive  cell of  the simple
cubic lattice).  As the area  of the disks is increased, the partition
function of this constrained model  becomes ever closer to that of the
unconstrained model. This  model with disks of the same  size is not a
glass:  in   fact  it  undergoes   an  Ising-like   phase  transition
\cite{Nelson}  to  a  state  which  is  one  of  the  two  differently
orientated   slightly    disordered   crystal   lattices    shown   in
Fig. \ref{16x16Disks083}.  In order  to investigate glassy behavior we
introduce in Sec.  \ref{binSOCM} a variant of the SOC model.  This has
two  species of  particles,  A and  B,  present in  equal numbers  but
randomly   distributed   over   the   plaquettes   as   indicated   in
Fig. \ref{spinmapping}.

Fig. \ref{spinmapping} also  shows that the SOC model  can be regarded
as a  spin model.   An effective spin  Hamiltonian is derived  in Sec.
\ref{spinH}  for  particles A  and  B  which  interact with  a  smooth
potential  $V(r)$,  e.g.    the  Lennard-Jones  potential.   For  such
potentials  it is possible  to calculate  analytically a  good leading
order effective spin Hamiltonian.  The Hamiltonian is very familiar in
the  field of  random  magnetic  systems; its  vector  spins have  $d$
components and interact with a $d$-component vector random field.  The
spin  interactions  are  a  mixture  of  exchange  and  pseudo-dipolar
couplings and there are also  single ion anisotropy terms.  Because it
is   so   well  understood   we   shall   just   briefly  outline   in
Sec.  \ref{SGbehavior} the phases  which can  exist for  the effective
spin Hamiltonian.   There are  choices for the  interatomic potentials
for which  the spin  Hamiltonian is in  the universality class  of the
Ising spin glass in a field and it this choice which is appropriate if
one is interested in the  properties of supercooled liquids or glasses
\cite{Yeo}.

The Ising  spin glass in a field  does not have a  phase transition in
dimensions $d  \le 6$ \cite{Moore12, AT6}, but  the correlation length
can become  large as  the temperature is  reduced, provided  the ratio
$h/J$ of the  standard deviation $h$ of the  random  field to
the  standard deviation $J$ of the   spin-spin  coupling  is
small.  In fact, we believe that for hard disks and spheres within the
SOC model  this ratio becomes  zero as the packing  fraction (density)
approaches  its maximum possible  value $\phi_{\text{max}}$.   The SOC
model  should therefore  show features  usually associated  with ideal
glass  behavior  in  this limit.  Much  of  the  paper is  devoted  to
investigating this intriguing possibility.

 Our  analytical  approach to  the  derivation  of  an effective  spin
 Hamiltonian does not extend to non-continuous potentials such as that
 appropriate to hard disks or spheres.  In order to study them we have
 had to resort  to simulations of the SOC  model, in particular, event
 driven  molecular dynamics.   The details  of this  are  described in
 Sec. \ref{methods}.

We  used  the   Lubachevsky-Stillinger  (LS)  algorithm  \cite{LS}  in
Sec. \ref{jam} to find some of  the jammed states of the SOC model for
hard  disks.   Its   jammed  states  are  similar  to   those  of  the
unconstrained  model. At the  densest packing  possible, $\phi_{\text{max}}$,
the state  is jammed. We obtain  an estimate of  $\phi_{\text{max}}$ from the
largest value  of the packing  fraction $\phi_J$ of the  jammed states
which we have found in small systems, for which there is a chance that
the LS algorithm might actually find the densest state. It is actually
very hard to do good simulations  in the region of most interest, that
is when  $\phi \to \phi_{\text{max}}$, because the  constraints introduced by
the  cell  walls makes  the  dynamics even  slower  than  that of  the
unconstrained system.  In two dimensions,
$\phi_{\text{max}}$  turns out to be very close to estimates of the
 glass close packing density $\phi_G$, which is sometimes identified with the random close
packing density $\phi_{\text{rcp}}$ \cite{PZ}.

We  study  in  Sec.   \ref{randomfield} and  Sec.   \ref{Correlations}
correlation functions  of the hard  disk system in order  to determine
the  variance  of the  random  field $h^2$  and  the  variance of  the
spin-spin couplings $J^2$.  The physical  reason for the presence of a
random field is also elucidated in Sec. \ref{randomfield}. The form of
the effective Hamiltonian is very  similar to that obtained for smooth
potentials in Sec.  \ref{spinH}: that  is, it is a mixture of exchange
and   pseudo-dipolar   couplings.    Unfortunately  because   of   the
difficulties associated  with the long  relaxation times as  $\phi \to
\phi_{\text{max}}$ we  cannot get good numerical estimates  of how $h$
and $J$ vary  with packing fraction in that  limit. Fortunately we can
provide  an argument in  Sec. \ref{Maxbehavior}  that shows  $h/J \sim
(1-\phi/\phi_{\text{max}})$           as           $\phi           \to
\phi_{\text{max}}$. 

One can use  the droplet theory of spin glasses  \cite{McM, BM, FH} to
determine the growth of the correlation length $\xi$ from the ratio of
$h/J$.  According to the droplet picture, the correlation length $\xi$
can  be estimated  by  equating the  energy  that can  be gained  from
flipping the spins  in a region of size $\xi$ in  the random field, $h
\xi^{d/2}$,  to  the  domain  wall   energy  cost  of  doing  this,  $J
\xi^{\theta}$, so
\begin{equation}
\xi \sim \left(\frac{J}{h}\right)^{\frac{2}{d-2\theta}}, 
  \label{droplet}
\end{equation}
 which reduces for $h/J \sim
(1-\phi/\phi_{\text{max}})$ to 
\begin{equation}
\xi \sim \frac{1}{(1-\phi/\phi_{\text{max}})^{\nu}}, \hspace{0.5cm}
\nu= \frac{2}{d- 2\theta}.
\label{SGnu}
\end{equation}
 $\theta$ is the  domain-wall exponent for Ising spin  glasses in zero
field.   For $d  =2$, $\theta  \approx -0.287$  \cite{Carter}  so $\nu
\approx 0.78$ while for  $d=3$, $\theta \approx 0.24$ \cite{Boettcher}
and $\nu \approx 0.79$.  Behavior of a power law kind is also expected
in RFOT at  a packing fraction $\phi_K < \phi_G$.   The value of $\nu$
in  that approach  is dependent  on  whether or  not \lq\lq  wetting''
effects  are considered  necessary  \cite{BB}), but  the wetting  form
$\nu=2/d$ is not  very different from that of  Eq.  (\ref{SGnu}) in two
and three dimensions due to the  fact that in these dimensions
$\theta$ is small. However, in our  approach, we have not seen any 
evidence for the ideal glass transition at $\phi_K$. For us the divergence 
of the correlation length is associated with glass close packing and jamming.

Finally  in  Sec.  \ref{Discussion}  we  discuss  the key question;  which
features  of supercooled  liquids and  glasses  can the  SOC model  be
expected  to describe  correctly?  It  is  argued that  the SOC  model
should be good for understanding  some of the phenomena which exist on
time scales less than the alpha  relaxation time, as the caging of the
particles  on time  scales  less  than the  alpha  relaxation time  is
mimicked by  the trapping  of the  particles in the  cells in  the SOC
model.  The  dynamical  correlation   length  is  extracted  from  the
properties of  correlations at the  alpha timescale so we  expect that
the SOC model should at least give $\nu$ correctly.

\section{The Single Occupancy Cell Model} \label{SOCM}

Cell occupancy models have a long history in the study of phase
transitions in fluids and liquids \cite{Hoover:1966cf, Hoover:1968ux}.
In the past, they have been used to calculate the equation of state of
hard spheres at high density \cite{Salsburg:1962dd} or to place bounds
on derivatives of the free energy \cite{Fisher:1965br} or entropy \cite{Hoover:1968ux}.
The system is divided into cells of a chosen geometry and a 
constraint is applied which fixes the number of particle centers
found in each cell. We focus on the single  occupancy cell (SOC) model, where
each cell can contain at most one particle. We work in two dimensions,
although the model is easily generalized to higher dimensions. Fig. \ref{MonaSOCM}
shows a hard disk fluid with a single cell occupancy constraint using square
cells. The constraints mean that disk centers interact only
with cell walls and disk surfaces interact only with other disk surfaces.

We note  that   it  might   be  possible  to   realize  the   SOC  system
experimentally, at least in two  dimensions. The square cells could be
produced by a wire grid, and a post could be attached at the center of
each disk so  that while the circumference of the  disk can pass under
the wire grid, the post at the center cannot.

SOC models are useful to us because they make the introduction of a 
spin representation of the problem straightforward. A disadvantage
of using the cell constraint is that at low packing fractions
the behaviour of the system deviates significantly from the behaviour
of the unconstrained system. At low packing fractions
most of the collisions will be between disks and cell walls,
so the cell geometry dominates.
As the packing fraction is increased, more collisions occur
between disks and close to jamming, almost all collisions
will be between disks. The closer the packing fraction is to $\phi_{\text{max}}$,
the better an approximation the constrained model becomes to the unconstrained
model as the cell walls no longer dominate the dynamics.

Another pecularity of SOC models is the appearance of singularities in
thermodynamic properties.  This occurs  because of how the constraints
limit the size of clusters  that can form.  Without constraints, it is
possible to find all particles forming a single cluster at all packing
fractions. This is  not possible in the constrained  system.  As shown
by   Hoover   and  Alder   for   a   one-dimensional   hard  rod   SOC
model\cite{Hoover:1966cf}, it  is only possible to form  clusters of a
certain size above a certain packing fraction. For example at very low
packing fractions, the constraints mean that clusters can only contain
at most two  particles. As the packing fraction  is increased clusters
can contain three then four  particles. At the packing fractions where
it  becomes  possible  for  larger  clusters to  form,  the  partition
function changes its analytic form and this means that discontinuities
appear  in  thermodynamic  quantities  such as  $\partial^2 P/\partial V^2$.   These
packing fractions get closer together approaching $\phi_{\text{max}}$,
and   the  discontinuities   decrease  in   size,  meaning   that  the
shortcomings become less important. We expect similar behavior in two
dimensions, but  as each disk  has more nearest neighbors  the effect
will  be smaller.  In any  case,  it has  not been  noticeable in  our
simulation results.

\begin{figure}
\begin{center}

	\resizebox{67mm}{!}{\includegraphics{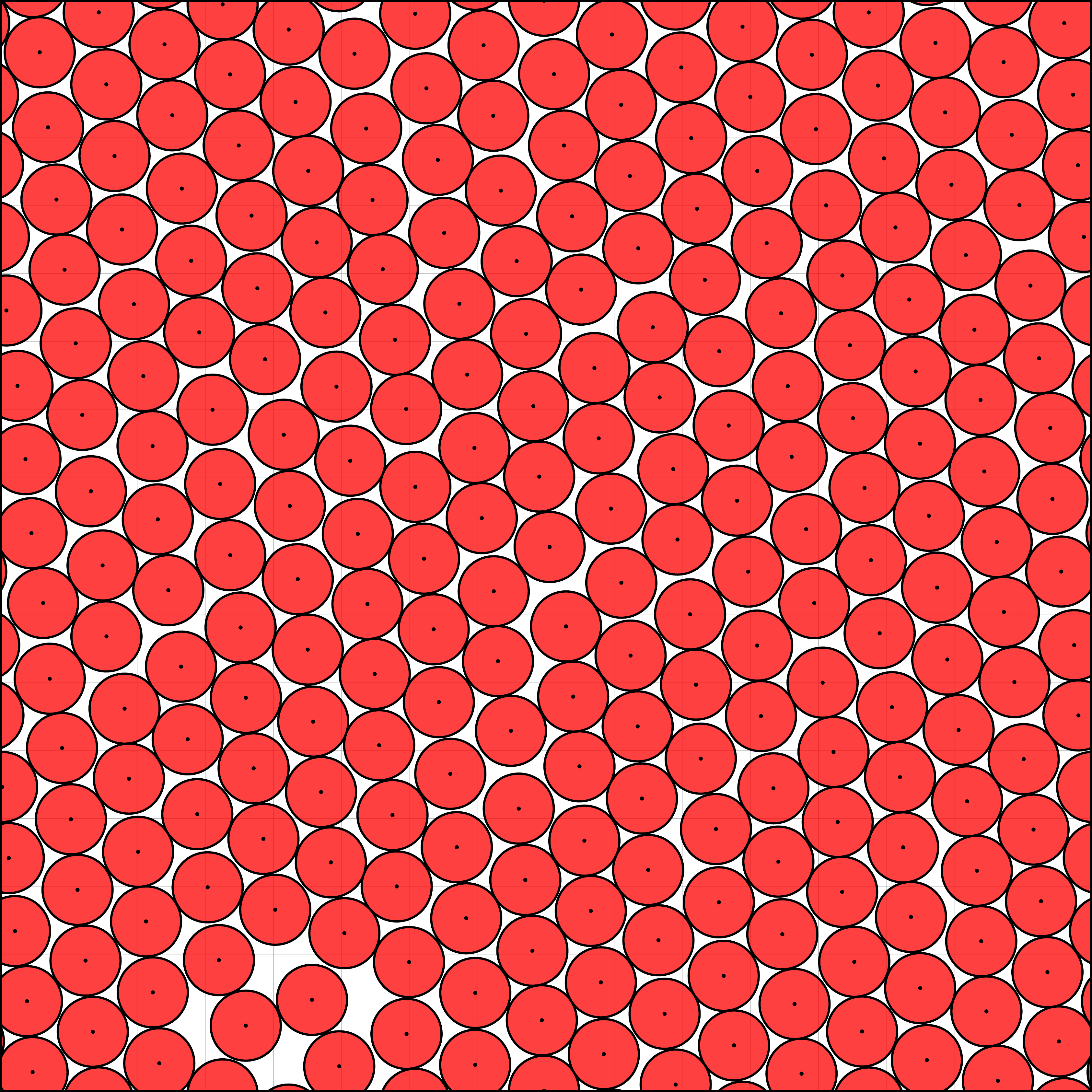}} \\
	 \resizebox{67mm}{!}{\includegraphics{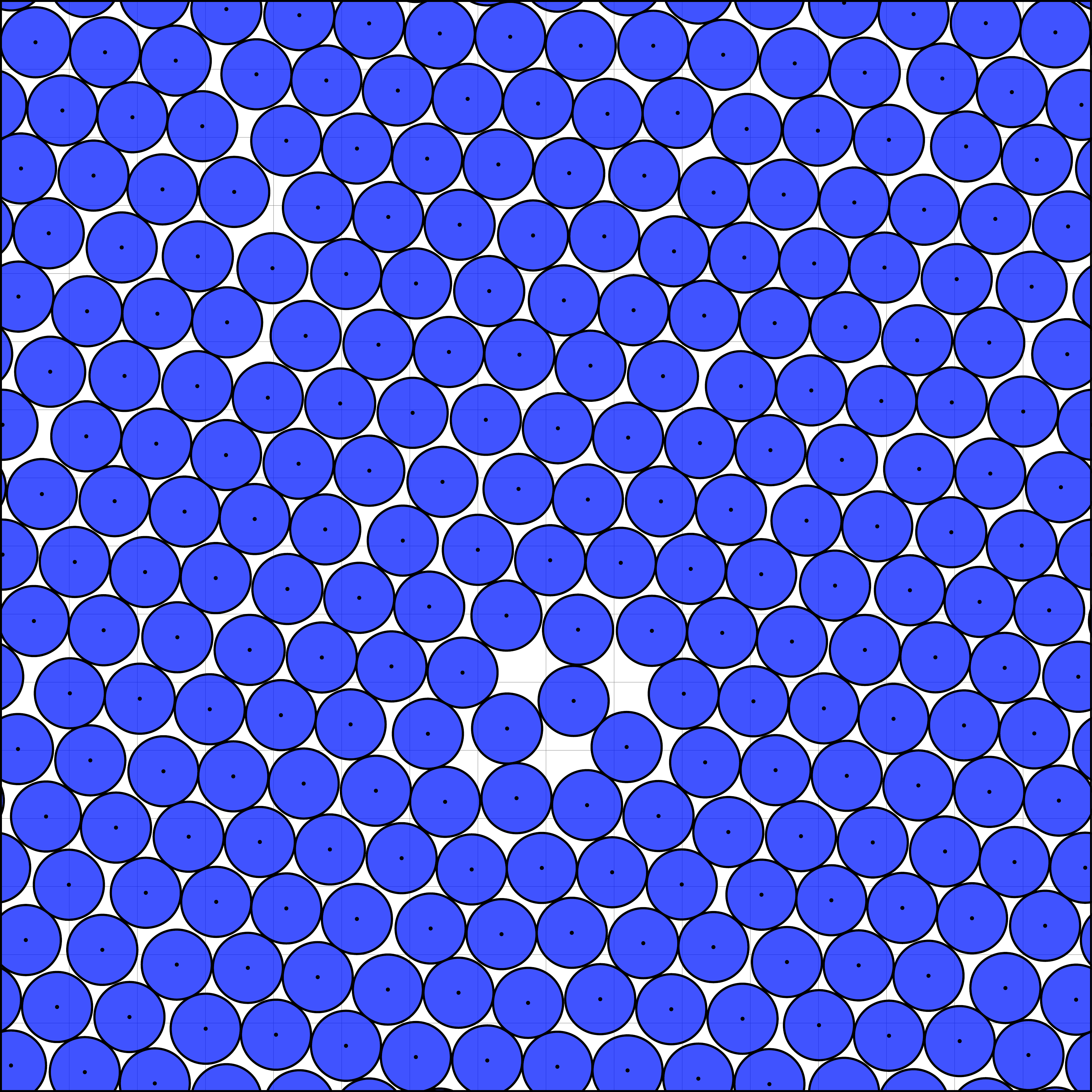}} 
	
	\caption[Two  possible orientations  at  high
          packing fraction]{ (Color online) A snapshot of two $16\times 16$ hard disk systems in the
          SOC  model at $\phi=0.83$,  each displaying  one of  the two
          possible                   defected                  crystal
          orientations.} \label{16x16Disks083}
\end{center}
\end{figure}

The model with  disks of the same size, as  in Fig. \ref{MonaSOCM}, is
not a glass. Without constraints the largest possible value of
the  packing  fraction  $\phi$ occurs when a triangular lattice with
all disks touching there neighbours is formed;  so $\phi_{\text{max}}
=  \pi/2 \sqrt{3} \approx  0.9069$. In the SOC version of the model, 
the constraints mean an exact triangular lattice cannot form, so we find
$\phi_{\text{max}} \approx 0.88$. When $\phi = \phi_c  \approx  0.77$ there  
is  a  phase  transition to  a  slightly disordered  crystal  which  is  
orientated  in  one  of  two  possible directions as in Fig.  
\ref{16x16Disks083}.  The critical exponents of this transition are expected to  
be those of the two-dimensional Ising model because  of this two-fold  
degeneracy of the orientation  of the slightly disordered crystal lattices \cite{Nelson}.

\section{Single occupancy cell models for modelling glasses} \label{binSOCM}

To make a glassy model, we introduce two different sizes of disk.
The binary disk system consists of hard disks of two species
(A and B), where the SOC constraints have been applied, as in Fig.   \ref{spinmapping}.  The  species  of  disk  have
different  radii  $\sigma_{\mathrm{A}}$  and  $\sigma_{\mathrm{B}}$, where  the  size  ratio
$R_{\mathrm{AB}} = \sigma_{\mathrm{A}}/\sigma_{\mathrm{B}}$ is held fixed as the packing fraction
is altered. The packing fraction is given by
\begin{equation}
\phi =\pi(\sigma_{\mathrm{A}}^2+ \sigma_{\mathrm{B}}^2)/2,
\label{defphibinary}
\end{equation}
where the side of the square plaquette has been taken to be of
unit length.  We set $R_{\mathrm{AB}}=1.0/1.4$ - this is a well explored choice
\cite{Teitel, Perera:1999hd, Hentschel:2007ig}. There are equal numbers
of each species ($N_{\mathrm{A}} = N_{\mathrm{B}} = N/2$), and each cell contains a disk
of species A or B with equal probability.

When $R_{\mathrm{AB}} \rightarrow 1$, the system undergoes crystallization
to   one   of  the   two disordered   crystal   states  shown   in
Fig. \ref{16x16Disks083} but with  substitutional disorder. Disks
of species $A$  and $B$ will be distributed  at random throughout the
defected crystal.

For $R_{\mathrm{AB}}=1.0/1.4$ without constraints, the densest
state is a phase separated crystal where the two species form separate
triangular crystals. Although this state is very stable, it takes such
a long time to form that it is rarely reproduced in simulations.
This makes the system a good model glass former. Recent work has shown that
phase separation may be achieved on simulational time scales in some
three dimensional binary systems \cite{Dyre}.
It may be that some nacscent phase separation could be driving behaviour
normally identified as glassy (slow dynamics, dynamic heterogeneity and
growing correlation lengths). For an example, see Ref. \cite{Hentschel:2008}.

With the introduction of the single occupany constraints, phase
separation can no longer occur as fixing the species of the disk in each
cell fixes the local composition of the hard disk fluid. We chose to 
distribute the species across the cells with equal probabilities. This
mimics what would happen if a well mixed fluid at low packing fraction
was rapidly `quenched' to a higher packing fraction without allowing
the disks to phase separate. As phase separation is prevented by the 
cell constraints, this means that
\textit{glassy behavior  can be  investigated in a  fully equilibrated
  model}; there  are no concerns that  if one runs  the simulation for
longer there will eventually be phase separation.

\section{The Spin Hamiltonian}\label{spinH}

\begin{figure}
\begin{center}
	\resizebox{80mm}{!}{\includegraphics{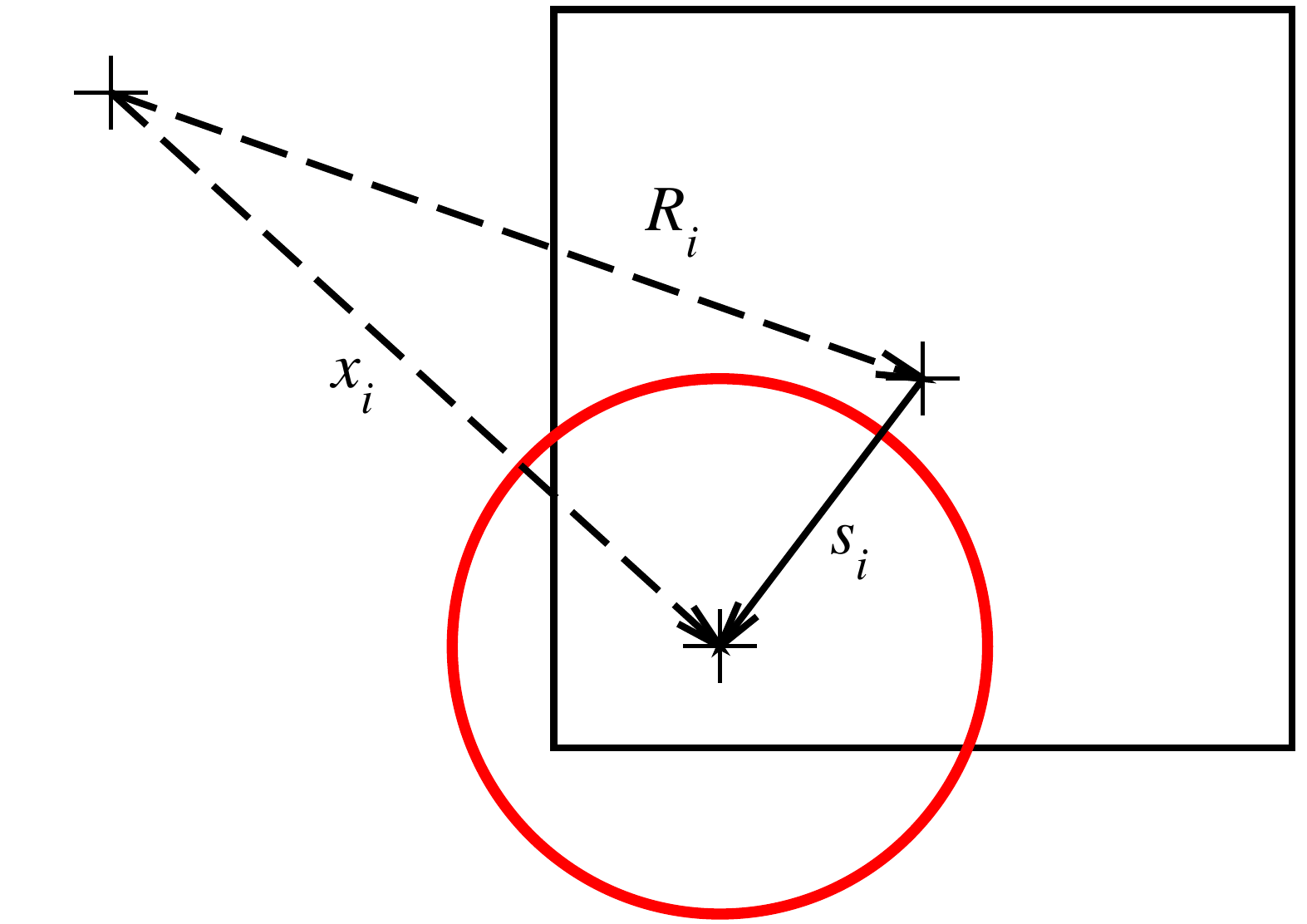}}
	\caption[The spin mapping]{ (Color online) Mapping  of a disk to a  spin -- the
          spin  $\vec{s}_i$ is  defined by  the equation  $\vec{x}_i =
          \vec{R}_i  + \vec{s}_i$  where $\vec{x}_i$  is  the position
          vector of disk $i$ and $\vec{R}_i$ is the position vector of
          the center of cell $i$. }\label{SM1}
\end{center}
\end{figure}

Our main reason for studying the SOC model is that it makes mapping to
a spin  system easy.  This is acheived by
drawing a  vector from the  center of each  cell to the center  of the
disk that  occupies that cell.  The  cells are labelled  $i$, where $i =1,2,
\ldots, N$ and the spin $\vec{s}_i$ is defined as
\begin{align}
\vec{x}_i = \vec{R}_i + \vec{s}_i,
\end{align}
where  $\vec{x}_i$  is  the  position  vector  of  the  disk  $i$  and
$\vec{R}_i$  is the position  vector for  the center  of cell $i$.
This mapping is illustrated in Figs. \ref{spinmapping} and \ref{SM1}.

The system may now be is analyzed as if it were a spin
system. Unlike the particles, the spins are equivalent and all
details of the interaction between them are found in the terms and
couplings of the spin Hamiltonian.
If we write $\vec{s}_i=(x_i,y_i)$, which is
appropriate for $d=2$, then in the usual XY model $|\vec{s}_i|=1$, but
in the SOC spin model, $x_i$  and $y_i$ take values which keep the disk in
the $i$th plaquette.

Let us suppose that the
particles in the  cells  interact with each  other through the
potential $V(r)$, where $r$ is the interparticle separation.
The hard  disk problem is a special  case of this potential
where $V(r)= \infty$ if $r$ is less than the sum of  the radii of the
two disks and is otherwise zero.  If we have a  binary mixture of two
types of  particles A  and B,  $V(r)$ will be  a shorthand for $V_{ij}(r)$,
where $i$ and $j$ encode the species of the interacting particles. We shall  now proceed to
derive  the  effective Hamiltonian  in  terms  of  the spin  variables
$\vec{s}_i$.

  Using the  notation in  Fig. \ref{SM1}, the  distance $r$  between a
  particle in cell $i$ and one in cell $j$ is
\begin{equation}
r =|\vec{x}_i-\vec{x}_j|=|\vec{R}_i+\vec{s}_i-\vec{R}_j-\vec{s}_j|.
\end{equation}
To second order in the spin variables,
\begin{eqnarray}
V(r)&=&V(R)+V'(R) \vec{R}\cdot(\vec{s}_i-\vec{s}_j)/R \notag \\
&+&\left[V''(R)-V'(R)/R\right]\left[\vec{R} \cdot (\vec{s}_i-\vec{s}_j)
\right]^2/(2 R^2) \notag \\
&+&V'(R) \left[\vec{s}_i-\vec{s}_j\right]^2/(2R) + \cdots,
\label{potential}
\end{eqnarray}
where $R=|\vec{R}_i-\vec{R}_j| \equiv R_{ij}$.

The partition function $Z$ of the SOC model  is 
\begin{equation}
Z=\int \prod_{i=1}^{N} dx_i dy_i \exp[-\beta \mathcal{H}],
\label{Zdef}
\end{equation}
where the integration  over ${x_i, y_i}$ covers the  area of the $i$th
plaquette.  The  Hamiltonian $\mathcal{H}$ is, to second  order in the
spin displacements, of the form (up to constants)
\begin{equation}
\mathcal{H}= -\sum_{i,\mu} h_i^{\mu} s_i^{\mu} 
-\frac{1}{2} \sum_{i,\mu,j,\nu} D_{ij}^{\mu \nu} s_i^{\mu}s_j^{\nu} + \cdots,
\label{hamiltonian}
\end{equation}
where the sums over $\mu$ and $\nu$  run from $1$ to $d$ and in $d=2$,
$\vec{s}_i=(x_i,y_i)$. The fields $h_i^{\mu}$ are given by
\begin{equation}
h_i^{\mu}= \sum_{j \ne i}V'(R_{ij})R_{ij}^{\mu}/R_{ij}.
\label{hdef}
\end{equation}
If all the particles are identical,  the \lq \lq
field'' term  $h_i^{\mu}$ is identically  zero. However, if we  have a
binary mixture of two types of particles A and B such that $V_{AA}$,
$V_{BB}$, and $V_{AB}$ all differ, then the field term  $h_i^{\mu}$ is non-zero
 and  time-reversal invariance
is broken. 

We can calculate the average  of $h_i^{\mu}$ when the average is taken
over  the   various  possibilities   allowed  by  the   selected  disk
distribution.  We will  consider just nearest-neighbor interactions to
illustrate  how the calculations  can proceed,  and the  case $\nu=x$.
Only the  sites to the right and left of  the site $i$ contribute
to the sum in Eq. (\ref{hdef}). At each of these sites there can be an
A or a B disk (with equal probability) and at the site $i$ there is an
equal probability of the disk being  A or B.  Summing over the various
possibilities  one finds $\overline{h_{i}^{\mu}}=0$.  The distribution
of     the     random     field     components    is     such     that
$\overline{h_i^{\mu}h_i^{\nu}}=h^2\delta_{\mu \nu}$, where
\begin{equation}
h^2=\frac{1}{4}[(V'_{AA}-V'_{AB})^2+(V'_{BB}-V'_{BB})^2].
\label{varh}
\end{equation}
The various derivative are calculated at the nearest-neighbor distance. Note that 
if $V_{AA}=V_{BB}=V_{AB}$, then $h=0$, as expected.

The quadratic term  in Eq. (\ref{hamiltonian}) takes the form for  $i \ne j$
\begin{align}
&H_{eff}(\vec{s}_i,\vec{s}_j)=-\frac{V'(R_{ij})}{R_{ij}} \times \notag \\
&\left[\vec{s}_i \cdot \vec{s}_j-
[1-\frac{R_{ij} V''(R_{ij})}
{V'(R_{ij})}](\vec{\hat{R}}_{ij} \cdot \vec{s}_i)(\vec{\hat{R}}_{ij} \cdot \vec{s}_j)\right], \label{coupling}
\end{align}
where the unit vector $\vec{\hat{R}}_{ij}$ is $\vec{R}_{ij}/R_{ij}$.
Note  if the interaction   $V(r)= -A/r^n$, Eq. (\ref{coupling}) reduces to
\begin{align}
H_{eff}(\vec{s}_i,\vec{s}_j)=-\frac{nA}{R_{ij}^{n+2}} 
\left[\vec{s}_i \cdot \vec{s}_j
-(n+2)(\vec{\hat{R}}_{ij} \cdot \vec{s}_i)(\vec{\hat{R}}_{ij} \cdot \vec{s}_j)\right] \notag
\end{align}
which for  $n=1$ is  the familiar dipole-dipole  coupling interaction.
For other non-power  law potentials $H_{eff}(\vec{s}_i,\vec{s}_j)$ can
be   regarded  as  a   mixture  of   the  exchange   interaction  with
pseudo-dipolar couplings. 

When  $i
\ne j$,  $D_{ij}^{\mu \nu}$ is of the form 
\begin{align}
D_{ij}^{\mu \nu}=\left[A_{ij}\delta_{\mu \nu}-
B_{ij} \hat{R}_{ij}^{\mu} \hat{R}_{ij}^{\nu}\right], 
\label{Ddef}
\end{align}
where 
\begin{equation}
A_{ij}=-\frac{V'(R_{ij})}{R_{ij}},
\label{Adef}
\end{equation}
and 
\begin{equation}
B_{ij}=-\frac{V'(R_{ij})}{R_{ij}}[1-\frac{R_{ij} V''(R_{ij})}
{V'(R_{ij})}].
\label{Bdef}
\end{equation}
For  $i =j$ there are single-ion anisotopy terms with coefficients
\begin{equation}
D_{ii}^{\mu \nu}=-\sum_{j\ne i} D_{ij}^{\mu \nu}.
\label{ion}
\end{equation}
For  smooth  potentials like  the  Lennard-Jones potential,  the
$h_i^{\mu}$   and  $D_{ij}^{\mu  \nu}$   can  therefore   be  directly
calculated.  The configurational  average and variance of $D_{ij}^{\mu
  \nu}$ due to  the quenched random distribution of A or B particles
in the  plaquettes can be obtained  by the method used  to obtain Eq.
(\ref{varh}); the expressions are complicated.

The only approximation which arises from the use of the Hamiltonian in
Eq.   (\ref{hamiltonian})  is  the   truncation  to  second  order  in
$s_i^{\mu}$. The hope  is that this truncation does  not alter the \lq
\lq universality class'' associated  with the phase transitions of the
spin system. Of course, further terms could be included if required.
 
When  Eq.  (\ref{ion}) is  used  to  fix  the single-site  terms,  the
Hamiltonian will  still have in its quadratic  terms the translational
invariance of Eq. (\ref{potential}).  Similarly if
\begin{equation}
h_i^{\mu}= \sum_{j \ne i} C_{ij} \hat{R}_{ij}^{\mu},
\label{heff}
\end{equation}
that will ensure  translational invariance in the linear  term 
in  Eq.  (\ref{potential}).   The  quantities  $A_{ij}$,  $B_{ij}$  and
$C_{ij}$ thus  specify an effective spin  Hamiltonian for our problem.

\begin{figure}
\begin{center}
	\resizebox{80mm}{!}{\includegraphics{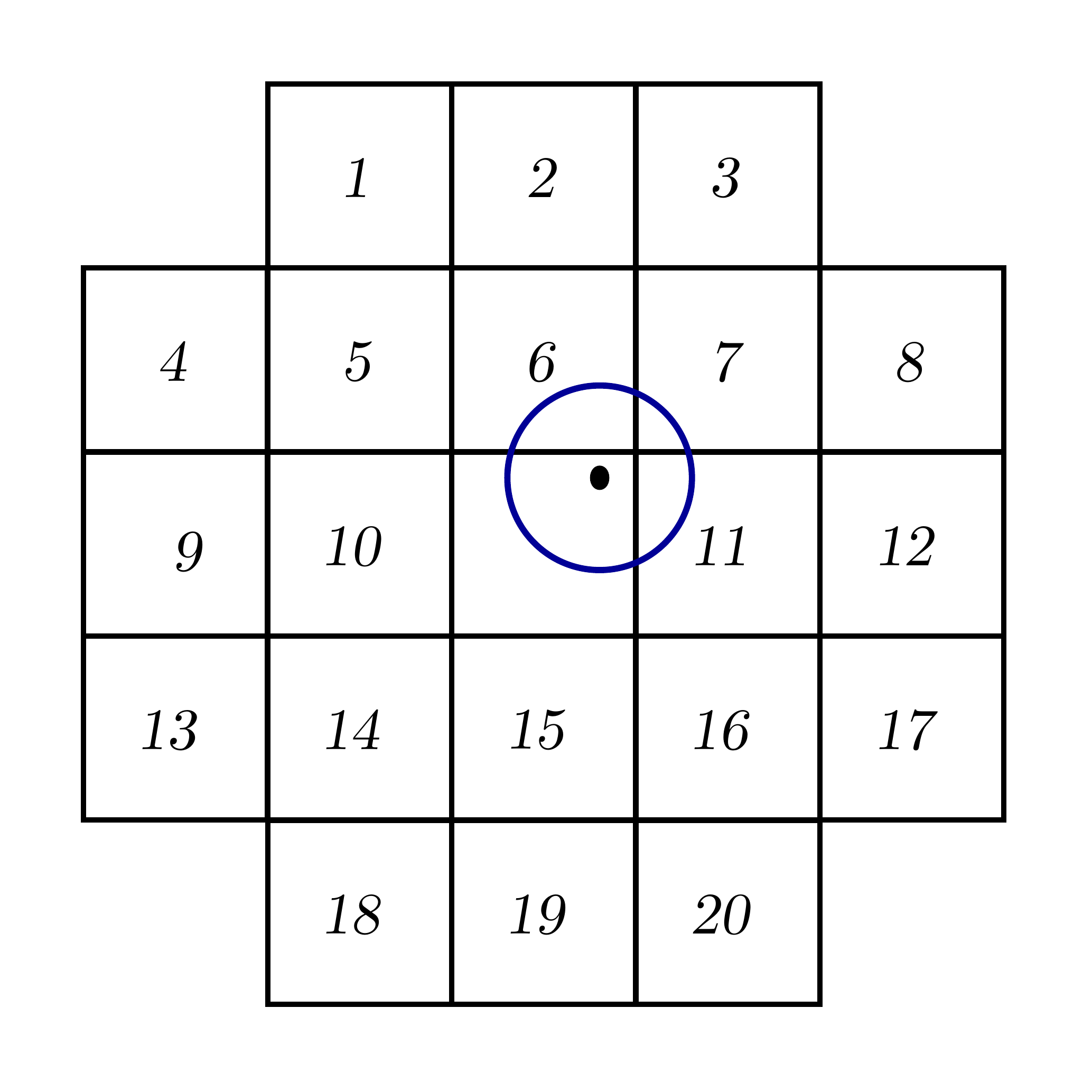}}
	\caption[Possible collision partners under the SOC]{ (Color online) Cells 1 to 20 contain disks which can possible collide with the blue disk when the hard disk fluid is at high packing fraction.}\label{LocEnv1}
\end{center}
\end{figure}

In Fig. \ref{LocEnv1} the positions  are shown of the plaquettes whose
associated disk  can interact with  the disk in the  central plaquette
when  the  packing  fraction is  high.  The  number  of such  disks  is
surprisingly  large; 20. At  smaller packing  fractions the  number is
reduced to 8.  (In  three dimensions the number at  large packing fractions
is 80). Now for the blue disk to interact  with the disk in  plaquette 1, the
disks in 2,  5, 6, and 10 must be occupying  only a restricted portion
of  their plaquettes.   A complicated  many-spin set  of terms  in the
effective spin Hamiltonian is needed  to describe this feature.  It is
clear  that  keeping   for  example  only  nearest-neighbor  spin-spin
interactions  does not  contain the  physics  of the  increase in  the
effective  number  of  interacting   spins  as  the  packing  fraction
increases. Truncating  the effective  Hamiltonian to just  binary spin
interactions  may also  fail  to capture  the properties  successfully
modelled by $p$-spin  models such as the dynamic  transition.  In this
paper, our  main concern  is the behavior  of glasses  at temperatures
below  the  dynamic  transition  temperature  or  at  densities  above
$\phi_d$, the  packing fraction associated with  the (avoided) dynamic
transition  (see Sec. \ref{Discussion})  and binary  spin interactions
are quite  sufficient to capture  the Ising spin glass  behavior which
prevails there.  An investigation as to whether the considerable range
of the spin  interactions can explain the utility  of mean-field ideas
in glasses \cite{PZ} is being carried out \cite{Godfreystatic}.

For hard disks  and spheres the potential $V(r)$  is infinite when $r$
is such that they overlap,  and zero otherwise. Such a potential makes
$V'(r)$ zero except at the  contact distance where it is infinite.  As
a  consequence the  expansion  used in  Eq.  (\ref{potential}) is  not
useful.  For hard disks and spheres we shall still use quantities like
$h_i^{\mu}$ and $D_{ij}^{\mu \nu}$,  but instead of deriving them from
the potential we will obtain their values as fitting parameters chosen
to  reproduce  measured correlations  (like  $\langle s_i^{x}  \rangle
\equiv \langle x_i \rangle$ etc.) rather  in the spirit of Ref. \cite{flocks}.
This is done in  Sec. \ref{Correlations}.

We have  already noted  that when $R_{AB}  \rightarrow 1$,  the system
will  undergo crystallization  to one  of the  two  disordered crystal
states in  Fig. \ref{16x16Disks083} but  with substitutional disorder.
The  disks  of species  $A$  and $B$  will  be  distributed at  random
throughout the  defected crystal. However,  when $R_{AB}$ is  close to
$1$ there  will be  effectively random fields  arising from  the small
differences in the A and B particles. We suspect that this changes the
transition  to  the  disordered  crystalline  state  to  that  of  the
random-field Ising  universality class. We  shall suppose from  now on
that   $R_{AB}$  is   sufficiently  different   from  unity   that  this
crystal-like transition no longer  arises and that only glass ordering
behavior  (i.e. spin-glass ordering in the spin mapping) need be considered.

\section{Spin Glass Behavior}\label{SGbehavior}

In this section we shall  discuss the properties of a spin Hamiltonian
like that  in Eq. (\ref{hamiltonian}).  For any smooth  potential, the
$h_i^{\mu}$ and $D_{ij}^{\mu \nu}$ can be directly calculated from the
potential. These  expressions will be  renormalized by the  effects of
multi-spin  interactions  neglected  in Eq.  (\ref{hamiltonian}),  but
hopefully they provide a good first approximation. For hard spheres or
disks  they  are  parameters  obtained  by  fitting  to  the  measured
correlation functions (see Sec. \ref{Correlations}).

For the binary SOC model, the $D_{ij}^{\mu \nu}$ between sites $i$ and
$j$ will depend on whether the  particles in the plaquettes are $A$ or
$B$ particles.   As the particles  can never escape from  their cells,
there is  quenched disorder present.   One can obtain  the probability
distribution function  of the  $h_i^{\mu}$ and $D_{ij}^{\mu  \nu}$ and
obtain their mean and variance.  Rather than do this, (which is rather
cumbersome  and  uninformative), we  will  just  outline  some of  the
possibilities which  might arise.  What  actually happens for  a given
set   of  potentials  $V_{AA},   V_{AB},  V_{BB}$   requires  explicit
calculations  and simulations  and the  number of  possible  phases is
large.  To limit the discussion  it is useful to recall the underlying
system: disks  (or spheres) whose  centers are trapped in  the squares
(cubes) of a square  (simple cubic) lattice. Ferromagnetic ordering in
the  spin system  would correspond  to a  crystallization of  the disk
centers into a  square lattice of the same periodicity  as that of the
plaquettes.   This  will not  happen if one uses an
appropriate  binary mixture for modelling glasses  
 and  so  we will  discount  the
possibility  of  a  transition  to  a  ferromagnetic  state  and  just
concentrate on  situations which are spin-glass like,  i.e. those where
the standard  deviation of the couplings  $D_{ij}^{\mu \nu}$ dominates
their mean  values. We  will therefore not  be discussing the  type of
ordering shown  in Fig.  \ref{16x16Disks083} (for the  monatomic system)
where there is clear crystal  order present: Glass behavior is not 
associated with any kind of long-range crystalline order.

The spins in the system are $d$-component spins so that one might have
thought  that  any  spin  glass   phase  in  this  system  would be  in  the
universality class  of the $d$-component vector  spin glass.  However,
it was shown  a long time ago \cite{BrayMoore82}  that in the presence
of pseudo-dipolar-like terms, the transition to the spin glass phase is
changed from  one in the  $d$-vector spin glass universality  class to
one in the Ising spin glass universality class.

The  random field  terms $h_i^{\mu}$  have  a dramatic  effect on  the
nature of the spin glass.  At mean-field level and for dimensions $d >
6$ a $d$-component random field present in a $d$-component vector spin
glass  produces   a  phase  transition  --   the  de  Almeida-Thouless
transition \cite{AT} -- which is  in the same universality class as an
Ising model  in a field \cite{SharmaYoung, Moore12}.   The presence of
the  pseudo-dipolar terms  just reinforces  the Ising  nature  of this
transition. For $d \le 6$ the  spin glass transition is removed by the
presence of the random field \cite{AT6, Moore12}.

 The  spin-glass  correlation  length,  which  is  equivalent  to  the
 point-to-set length  scale, can still become  large for $d  \le 6$ if
 the ratio $h/J$ is small. ($J$ is a measure of the standard deviation
 associated with  the $D_{ij}^{\mu  \nu}$).  According to  the droplet
 picture \cite{McM,  BM, FH} the  correlation length $\xi$  depends on
 this ratio as in Eq. (\ref{droplet}).  This is the correlation length
 appropriate to $T = 0$.  As a function of temperature the correlation
 length is small  at high temperatures and grows to  this value in the
 limit when  $T \to 0$.  We  expect that $\xi$ might  become large for
 real fragile  glasses at low temperatures.  However,  on this picture
 $\xi$  will never  become infinite  unless  $h/J$ goes  to zero.   We
 suspect  that this  never happens  for smooth  potentials.   In other
 words, for such  potentials no diverging length scale  is expected in
 $d \le 6$.

Note that if we had used the mean-field approximation to determine the
properties of the spin system, we would have found a phase transition,
the de  Almeida-Thouless transition,  at a finite  temperature provided
the ratio $h/J$  is not too large. We would have  then been tempted to
identify this transition with  the ideal glass transition. However, it
is our belief that the AT  transition does not occur for dimensions $d
\le 6$ \cite{Moore12,AT6}.

One  might further  wonder  whether the  multi-spin  \lq \lq  p-spin''
interactions  which  were  alluded   to  in  the  discussion  of  Fig.
\ref{LocEnv1} might  make a transition to a  one-step replica symmetry
broken  state possible.   We have  neglected them  in  our discussion.
This is  the scenario envisaged in the  RFOT and is the  origin of the
ideal glass transition.   We do not think such  a transition can exist
outside the mean-field approximation, that is, in finite dimensions, where
the one  step replica  symmetry broken state  is unstable  against the
thermal excitation of large droplets \cite{Moore06}.

For binary mixtures of hard spheres and disks, a mechanism might exist
to drive the ratio $h/J$ to zero.  In the SOC model there is a maximum
packing fraction for hard disks or spheres.  For our binary mixture of
hard disks,  this value  is estimated in  Sec.  \ref{jam}.   Its value
$\phi_{\text{max}}$  is  very similar  to  $\phi_{\text{rcp}}$ of  the
unconstrained  model  and  in  both  models at  these  densities,  the
pressure  is  infinite.   We  shall  present  numerical  evidence  and
arguments  in  Sec.   \ref{Correlations}  that  the  ratio  $h/J  \sim
(1-\phi/\phi_{\text{max}})$, so that in  our version of the SOC model,
the  correlation  length  $\xi$  diverges  to  infinity  according  to
Eq. (\ref{droplet}).   In other  words there are  features of  a glass
transition in the hard disk SOC model as $\phi \to \phi_{\text{max}}$,
in that  there is  a diverging correlation  length in spin--glass--like
correlation functions.  The rest of this paper is devoted to the study
of this behavior.

To acquire data  to determine $\phi_{\text{max}}$ and to  obtain estimates of
$h$ and $J$,  it is necessary to perform simulations  of the SOC hard disk
system.  In the  next  section,  our simulation methods for hard disks
 are outlined.

\section{Event driven molecular dynamics and the Lubachevsky-Stillinger algorithm} \label{methods}

To simulate the hard disk system, we use event driven molecular dynamics
following the method described by Lubachevsky \cite{Lubachevsky:829677}.
This is an efficient way to perform simulations of hard disk systems.
We will not describe the method in full here, but the basic principle involved
is to keep a list of the next collision each particle will be involved in 
ordered by time. Time is moved forward by jumping to the collision that 
occurs next, and then recalculating the list in light of the new velocities
and positions the colliding particles now have. The speed of the simulation
is further boosted by the fact the cell constraints restrict the particles
that can possibly collide.

We generate configurations at a particular packing fraction by first placing
particles randomly in each cell in such a way that each cell is equally
likely to contain a particle of either species. The particles start with
zero radius (so there is no possibility of overlap) and at time $t$ have
radius $\sigma_i(t) = \Gamma_it$ where the $i$ denotes the species of 
the particle in question. The growth rate $\Gamma_i$ is set to be small
to allow the disks to remain in equilibrium as the simulation progresses.
We use $\Gamma \sim 10^{-4}$.
Each disk is given a random velocity so that $|\vec{v}_i|=1$

The disks are allowed to collide and grow
until the system reaches the desired packing fraction. Then
the disk radii are set to be constant and measurements may be made.

To generate  jammed states, we  make use of  the Lubachesky-Stillinger
algorithm \cite{LS}. We begin the  simulation as described above, but in
this case  the growth  rate of  the disks is  not set  to zero  at any
time. As the simulation proceeds, collisions (events) become separated
by smaller and  smaller time intervals and the  simulation will become
slower. If  $\delta t_{DD}$ is the time  between disk-disk collisions,
$1/\delta  t_{DD}  \rightarrow \infty$  as  the simulation  proceeds.
This   is  equivalent   to   a  divergence   in   the  pressure.   The
Lubachevsky-Stillinger  algorithm   works  by  choosing   a  value  of
$\delta t_{DD}^*$ below  which collisions  are close enough  together that
the  system has  effectively jammed.  Here  we use  $\delta t_{DD}^*  =
10^{-8}$ with $\langle|v_i|\rangle=1$. Repeating the simulation
  yields a range
of  jammed  configurations,  with  a distribution  of  jammed  packing
fractions $\phi_{\mathrm{J}}$.

\section{Determining  the maximum packing fraction in the  SOC binary model} \label{jam}

In  this  section  we  shall  estimate the  largest  packing  fraction
$\phi_{\text{max}}$  for our  binary hard  disk system.  It is  as  $\phi \to
\phi_{\text{max}}$  that we  expect  the correlation  length  to diverge,  so
$\phi_{\text{max}}$ is like the critical temperature of the system.

As already  discussed, at low  packing fractions the behavior  of the
constrained  fluid is very  different from  that of  the unconstrained
fluid, becoming closer to it as the packing fraction is  increased. At some
packing  fraction the  system will  jam.  In  a jammed  state  for the
unconstrained  system,  the  disks  are  held in  place  by  their  $z$
neighbors, (except for a few  rattlers), where $z=2d$ -- the so-called
isotatic condition \cite{Torquato10}.  In the  SOC model a disk can be
jammed when its center is pinned against a plaquette wall.

\begin{figure}
\begin{center}
	\resizebox{85mm}{!}{\includegraphics[angle=0]{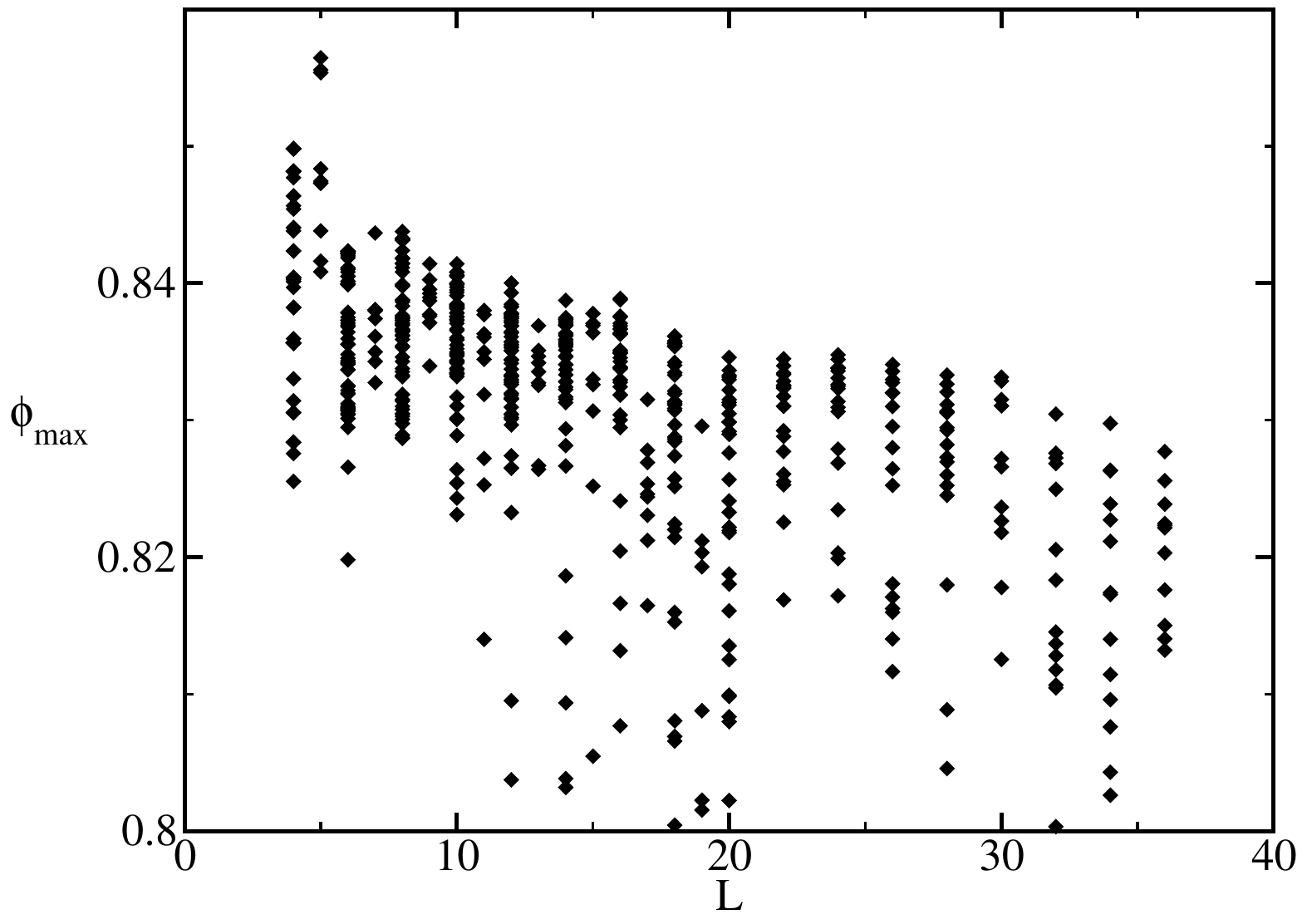}}	
	\caption{The jammed packing fractions  for the SOC binary hard
          disk fluid found  in  simulations from different
          random     starts    at     various     linear    dimensions
          $L$.} \label{MaxPhiBin}
\end{center}
\end{figure}

Jammed  states were obtained  for a  range of  system sizes  using the
Lubachevsky-Stillinger     algorithm     \cite{LS}    described     in
Sec. \ref{methods}.  A plot of the  values of $\phi_J$  values for the
binary  hard disk  system can  be seen  in Fig.   \ref{MaxPhiBin}. The
plot does not show the complete range of jammed states possible in the
$L \times L$, but a subset obtained from several runs of the algorithm.

  There is  a fall from  $\phi_{\text{max}} \approx 0.855$, when  $L=5$, then
  $\phi_{\text{max}}(L)$  settles around  $\phi =  0.835$ before  falling off
  slightly  when  $L>30  $.  This  fall  off is due  to
  inefficiencies in the simulation for generating jammed states of the
  highest packing densities.  The  value $\phi_{\text{max}} \approx 0.835$ is
  quite  close  to  the  value   of  the  packing  fraction  at  which
  unconstrained    binary   disk   systems    of   this    type   jam,
  $\phi_J\approx0.84$,     using      the     protocol     studied     in
  Ref.  \cite{Teitel}. In  other words,  it  is close  to the  numbers
  quoted  for  \lq  \lq random  close packing''  in  two  dimensional
  systems.

The most significant change from the unconstrained binary fluid is the
presence   of   a   \textit{well-defined}  maximum   jamming   density
$\phi_{\text{max}}$.   In  the   unconstrained  fluid  when  a  jammed
configuration  has been  acquired,
one can always imagine  creating a denser state by rearranging
a  few  of the  particles  into a  region with more local crystalline order.  
This  will create a small amount  of free volume
which will allow  further arrangements to be made.   If this programme
is   continued,  the   final   point  is   a  completely   crystalline
configuration.   A continuum  of states  at packing  fractions between
$\phi_{J}$  and $\phi_{crystal}$  can be  constructed by  this method 
(although there is no guarantee that they will be stable).
This  makes defining  a densest  non-crystalline  state problematic.
However, since  the cell constraints  do not allow the  composition of
the fluid to be altered, this  programme cannot be followed in the SOC
model  and  there is  indeed  a  well-defined  maximum density.   This
maximum  density will  depend  on the  particular  realization of  the
distribution  of large  and  small disks  over  the cells,  but it  is
probably a self-averaging quantity.

In the  unconstrained model each protocol for  producing jammed states
produces  states with  a characteristic  value of  $\phi_J$ as  $N \to
\infty$.  The  Lubachevsky-Stillinger  algorithm  used in  this  paper
produces, in the SOC model, states of a characteristic $\phi_J$, which
will  not  in  general  include  the states  at  $\phi_{\text{max}}$,  except
possibly at small values of  $N$.  Within the SOC, different protocols
will  also produce  different  values for  $\phi_J$.  Protocols  which
produce  jammed states  whose $\phi_J$  is close  to  $\phi_{\text{max}}$ are
producing jammed  states closer to  those in the  unconstrained model.
As a consequence, we are expecting that for the densest jammed states,
nearly all  the disks  will be  touching 4 other  disks in  the jammed
state and  very few, if  any, will be  jammed because their  center is
touching a plaquette wall. In principle, but probably not in practice,
one could obtain estimates of $\phi_{\text{max}}$ by calculating the pressure
$P$  in a fully  equilibrated system  and determining  $\phi_{\text{max}}$ by
fitting to
\begin{equation}
\frac{PV}{N k_B T} =\frac{d \phi}{\phi_{\text{max}}(1-\phi/\phi_{\text{max}})},
\label{Pdef}
\end{equation}
which becomes  exact as $\phi  \to \phi_{\text{max}}$ \cite{Salsburg:1962dd}.
The problem  with using this procedure  is that it  is  very  hard to
equilibrate the system at packing fractions close to $\phi_{\text{max}}$.

The cell  constraints affect the
dynamics of the system. This is a key concern as it affects how
quickly  the system  can  be  equilibrated and  hence  the quality  of
simulations  which  can be  done.   The  system  is clearly  glassy  -
simulations  performed   on  systems  with   packing  fractions  above
$\phi=0.75$  become  noticeably slow,  while  approaching the  maximum
packing  fraction  of  around  $\phi \approx 0.835$  makes  good  measurements
extremely hard. \textit{The presence of the cell constraints
  makes  the  dynamics even  slower  than  that  of the  unconstrained
  system}.

Imagine  a binary  fluid at  high  packing fraction,  focussing on  one
single disk.  At any given time there  will be a variety  of moves the
disk will be  able to make. Most will be short  and rapid (the typical
behavior  of  a  caged particle),  but  some  may  be part  of  large
rearrangements that will allow the structure of the fluid to relax and
change.  It is  reasonable to  assume that  the cell  constraints will
block a lot  of these movements (simply because the  walls of the cell
will  intercept the  paths the  disk wants  to take),  and they are more
likely  to  interfere  with  the  longer paths.  Thus  with  the  cell
constraints in place,  it is expected that the  dynamics of the system
will become  slower. Lots of local  rattling will be  allowed, but the
system  will  have  to  wait  for longer  before  large,  co-operative
movements that allow structural rearrangements  take place.

\section{Origin of the Random Field For Hard Disks}\label{randomfield}

\begin{figure}
\begin{center}
	\resizebox{80mm}{!}{\includegraphics{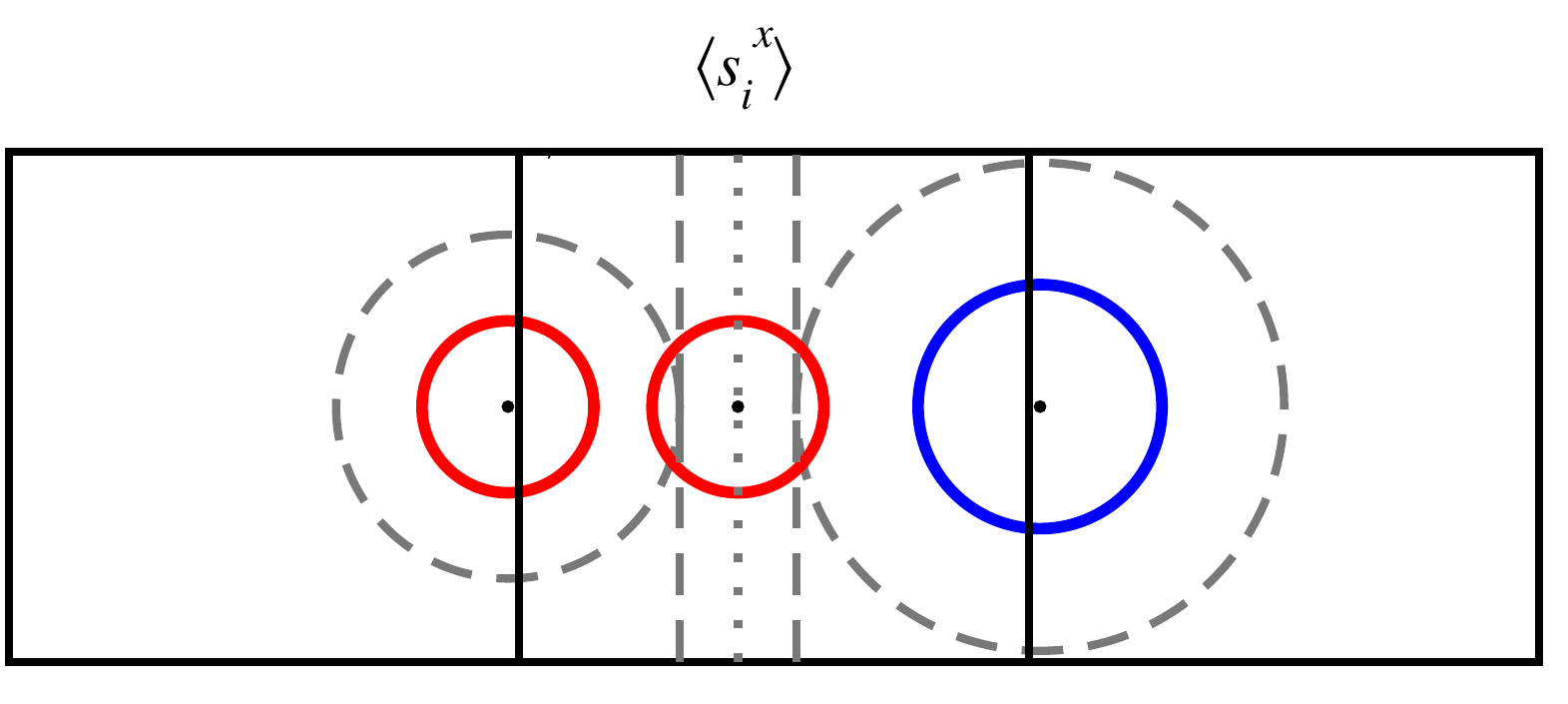}}	
	\caption{ (Color  online) Origin  of non-zero local  field and
          hence magnetization  $\langle s_i^{x}\rangle$ in  the binary
          disk system. Disks  of different size on either  side of the
          central disk  bias it  in one direction  - in this  case the
          larger disk  on the right  forces the central disk  to spend
          more time on the left of its cell.} \label{LocMag}
\end{center}
\end{figure}

We have already remarked that when all the particles are identical the
field term $h_i^{\mu}$ in  Eq. (\ref{hamiltonian}) is zero. For binary
mixtures  it is non-zero  and this  makes the  expection of  the local
magnetization $\langle s_i^{x} \rangle$  also non-zero. This is easily
understood from Fig. \ref{LocMag}.

\begin{figure}
\begin{center}
	\resizebox{80mm}{!}{\includegraphics{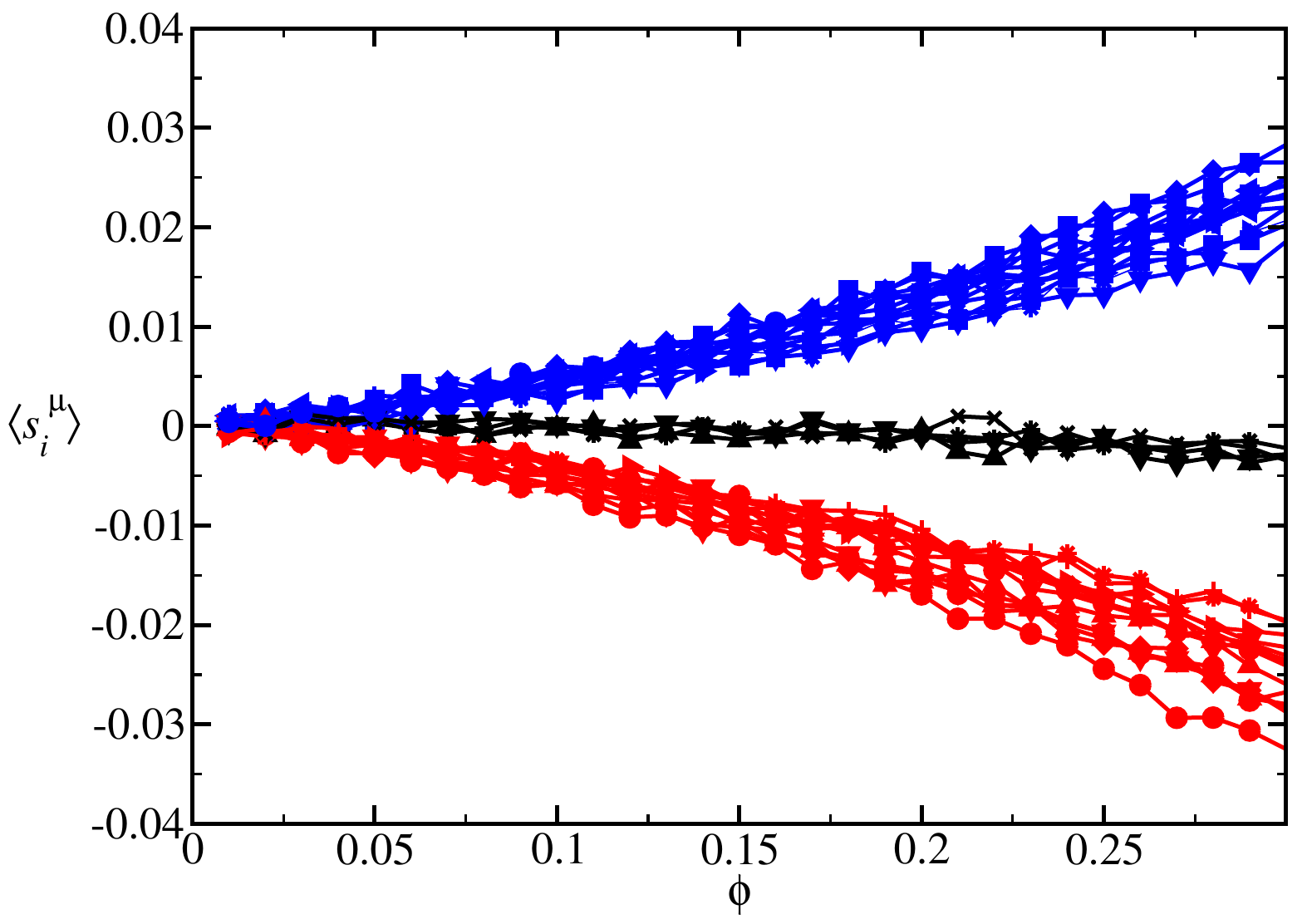}}	
	\caption[Local magnetization for  spin system derived from the
          binary  hard disk  fluid at  low packing  fractions]{ (Color
          online)  All  components  of  local  magnetization  $\langle
          s_i^{\mu}  \rangle$ for  the  spin system  derived from  the
          binary hard disk fluid under the SOC at low packing fraction
          for a  $4 \times  4$ system. Note  the three  distinct bands
          into which they fall.}
\label{LocMagLow}
\end{center}
\end{figure}

Zero  local   magnetisation  means  that   a  disk  spends   its  time
symmetrically distributed over its cell.  With this in mind it is easy
to see  why the  local magnetization  is finite at  all $\phi$  in the
binary system. When the packing  fraction is very low the disks rattle
backwards  and forwards  in their  cells, rarely  colliding  with each
other. The  finite local  magnetization is caused  by having  disks of
different sizes  on either side  of the central  disk. Say there  is a
large disk  to the right, and  a small disk  to the left (as  shown in
Fig.  \ref{LocMag}). The neighboring disks will intrude into the cell. 
When their  sizes are
different  they can intrude  by  different  amounts. In  the case  just
described the central disk will spend  more time on the left hand side
of  the cell  as there  is  more free  volume there.   As the  packing
fraction  is increased,  there is  more intrusion  by  the neighboring
disks and  the deviation  from the center  of the cell  becomes larger
This means that the local  magnetization gets  larger.  This
suggests that  there should be  three different types of  behavior for
the local magnetization: large disk to  the left and small disk to the
right ($\langle s_i^x \rangle >0$),  large disk to the right and small
disk to the left ($\langle s_i^x  \rangle <0$) and lastly disks of the
same size  on each side ($\langle  s_i^x \rangle \sim  0$). In Fig.
\ref{LocMagLow} the components of the local magnetisation split into these
three groups. At higher packing fractions, the groups blur into one due to
interactions between increasing numbers of disks, and the components
are randomly distributed about zero.

In  the spin  interpretation,  a finite  local magnetization  randomly
distributed about  zero implies the  presence of a local  random field
$\vec{h}_i$  interacting with  each spin  through a  term of  the form
$-\sum_i  \vec{h}_i.\vec{s}_i$.  The
expectation value of the  total magnetization $\langle \vec{M} \rangle
=0$,  where  $\vec{M}=\sum_i \vec{s_i}/N$,  for  all packing  fractions.
This suggests  that the  $h_i^{\mu}$   must  be  evenly
distributed around zero.  The source of the random nature of the field
is  the random distribution  of the  species of  disk over  the cells,
since this  affects the local  magnetization at all  packing fractions
through the  mechanism described above.  This field  will be discussed
again in the following sections.

\section{Spin-Spin Correlations}\label{Correlations}

We now study correlations of the form $\langle
s_i^{\nu}s_j^{\mu} \rangle$ where $i$  and $j$ index the lattice sites
and  $\nu$  and  $\mu$  label  the  $x$  and  $y$  components  of  the
spins. For a  spin in cell $i$, we can calculate $\langle
s_i^xs_j^x \rangle$, $\langle s_i^ys_j^y \rangle$, $\langle s_i^xs_j^y
\rangle$ and  $\langle s_i^ys_j^x \rangle$ for  nearest neighbors (the
spins  north,  south, east  and  west of spin $i$)  and
next-nearest neighbors  (the spins north-east,  south-east, south-west
and  north-west of  the spin  $i$).  We are interested in  using these correlations
as a  guide to the effective  interaction between the  hard disks. Our
studies suggest  that the  effective spin interactions  follow closely
the form expected in  Sec. \ref{spinH}: the effective spin Hamiltonian
is well-approximated by Eqs. (\ref{hamiltonian}) and (\ref{Ddef}).

\begin{figure}
\begin{center}
	\resizebox{80mm}{!}{\includegraphics{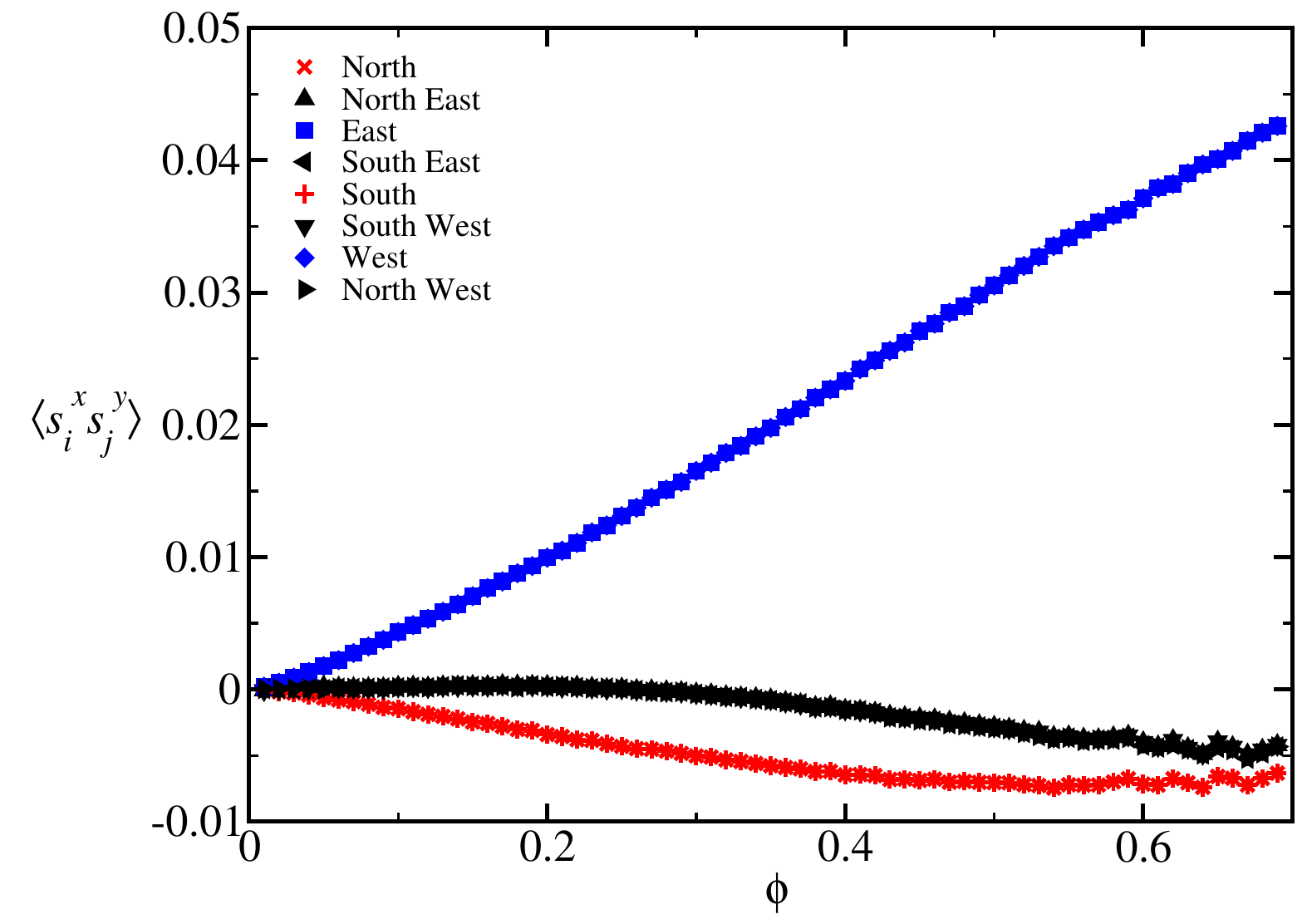}} \\
	\resizebox{80mm}{!}{\includegraphics{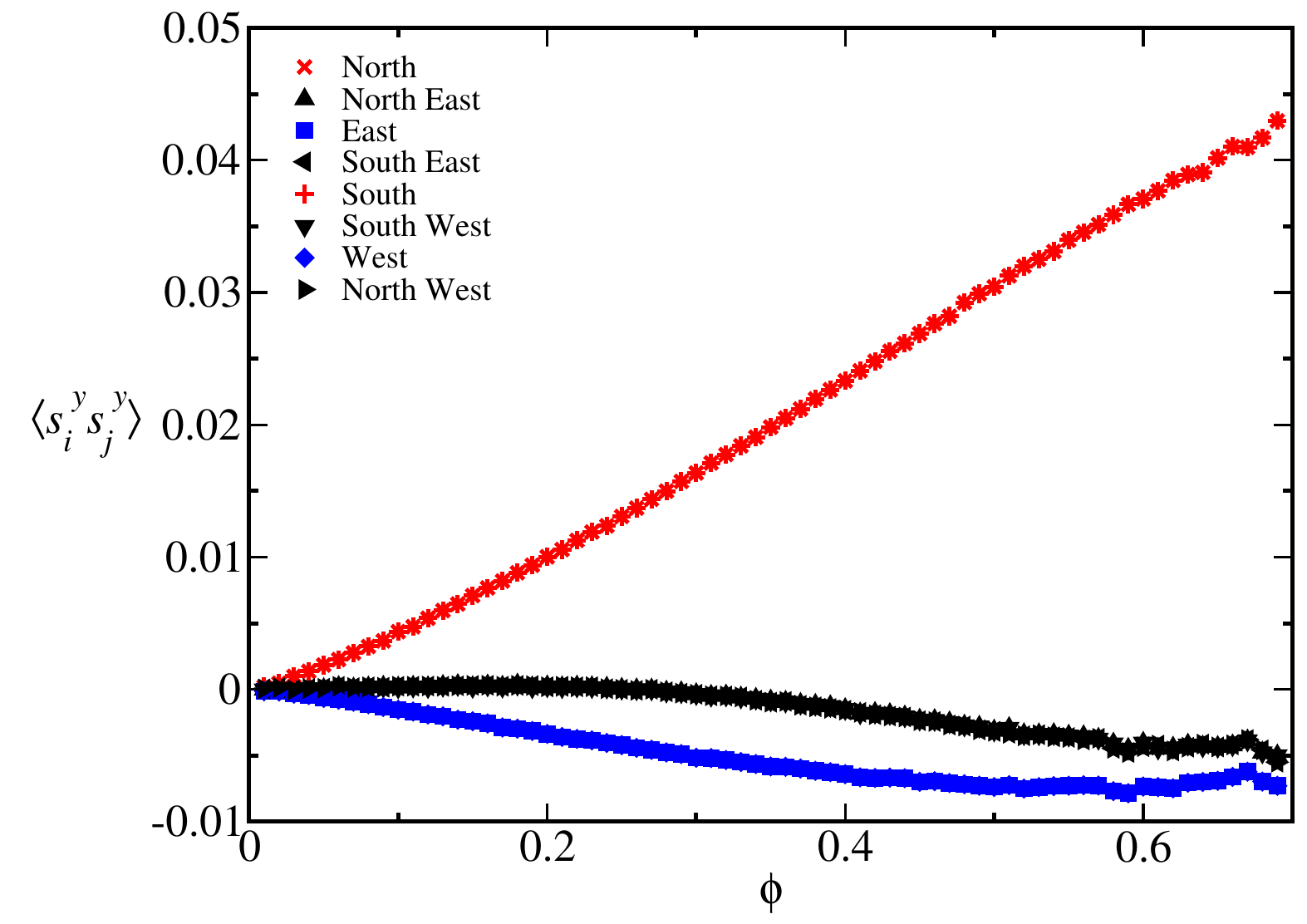}} 
      \caption[Spin-spin  correlations   $\langle  s^x_is^x_j  \rangle$  and
  $\langle s^y_is^y_j \rangle  $ for the spin system  derived from the
  binary hard  disk fluid under the SOC.]   { (Color online) Averaged spin-spin correlations 
    $\langle  s^x_is^x_j  \rangle$  and  $\langle  s^y_is^y_j
  \rangle$   between  spin   $i$  and   its  nearest   neighbors  and
  next-nearest neighbors, averaged over sites $i$ and disk realizations.  Note that sets of points with the same color all lie on top of each other.}
 \label{XXYYBin}
\end{center}
\end{figure}

\begin{figure}
\begin{center}
        \resizebox{80mm}{!}{\includegraphics{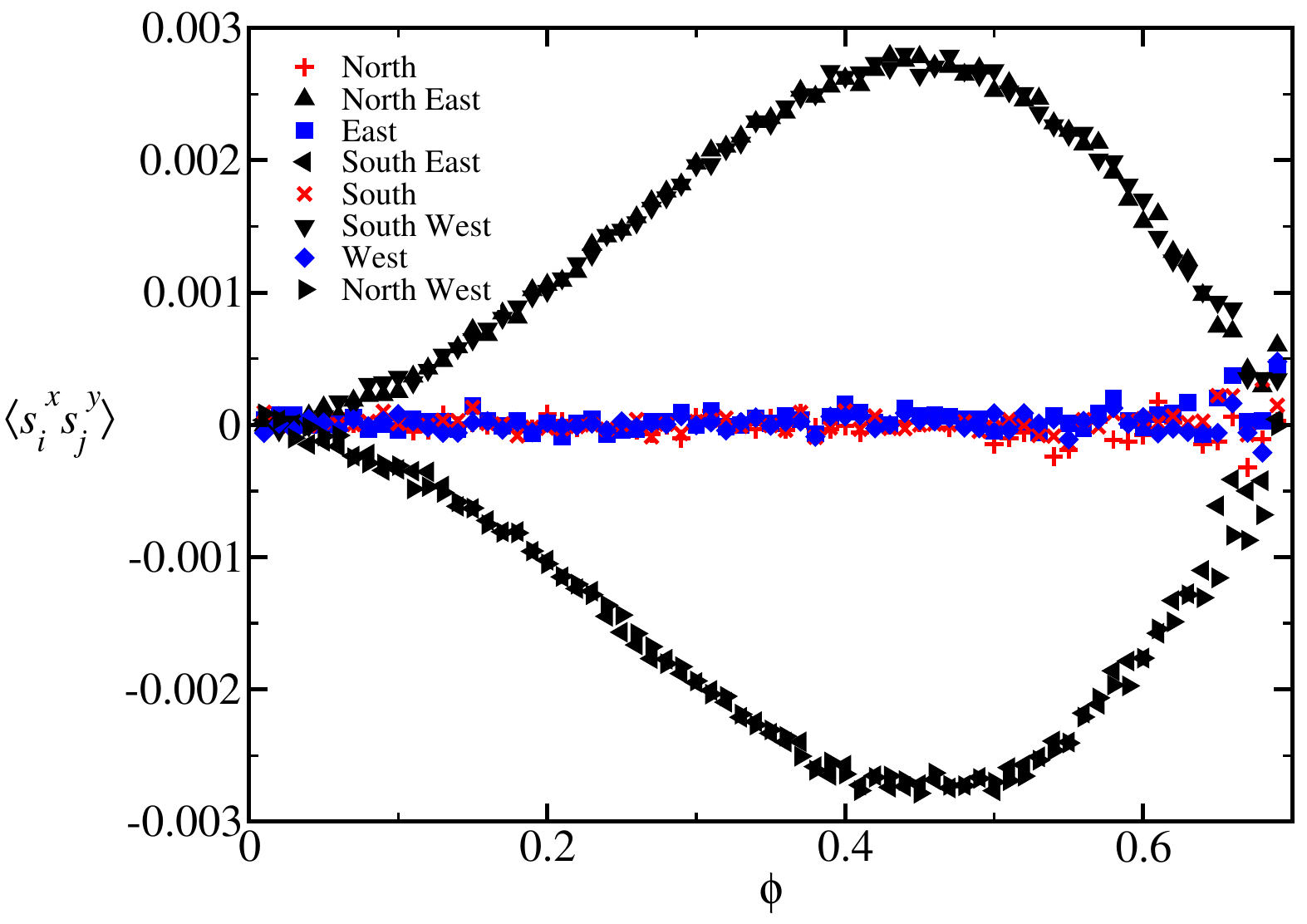}} \\
	\resizebox{80mm}{!}{\includegraphics{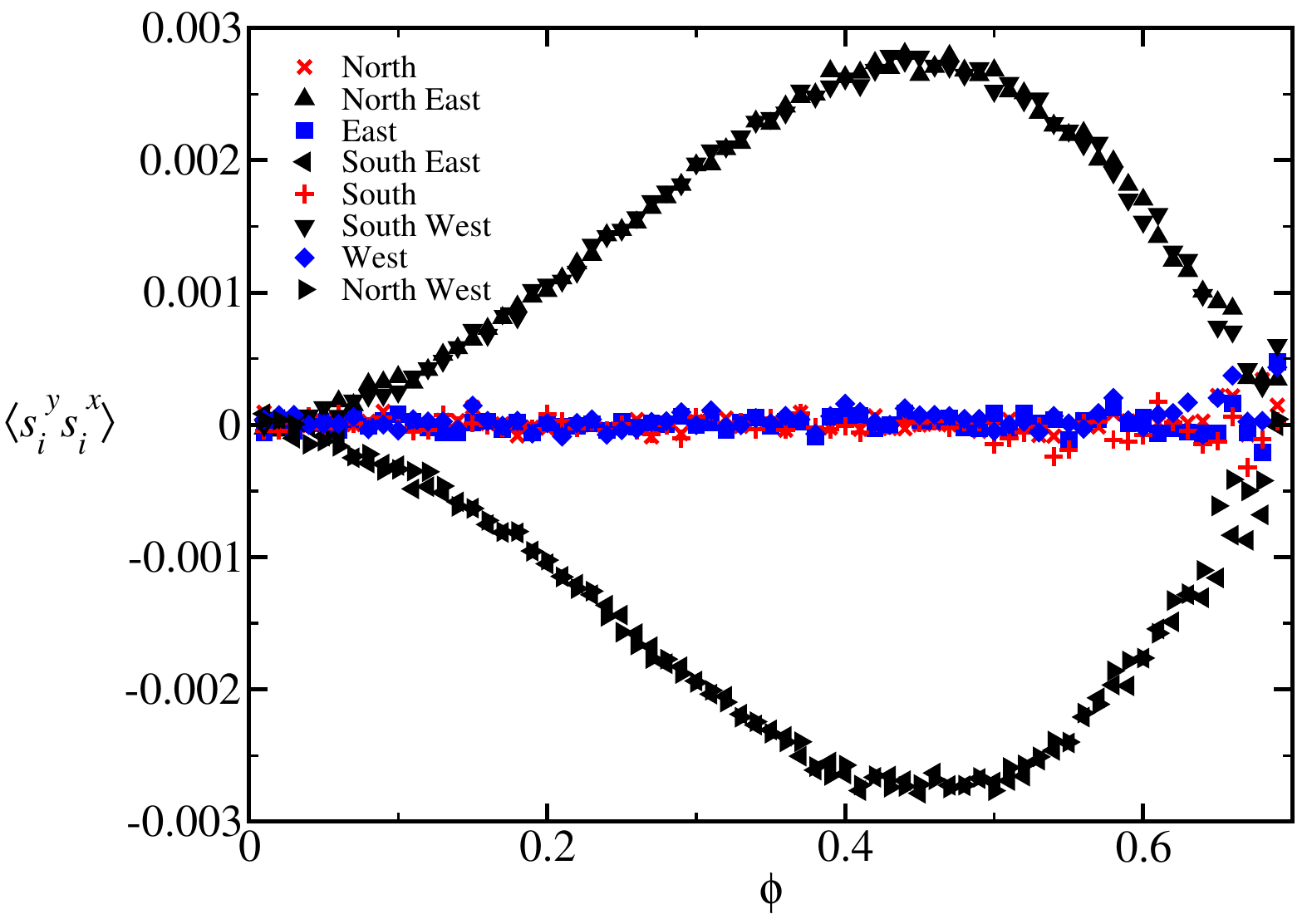}} 
		\caption[Spin-spin  correlations  $\langle  s^x_is^y_j
                  \rangle$  and $\langle  s^y_is^x_j \rangle$  for the
                  spin system derived from  the binary hard disk fluid
                  under the  SOC.]{ (Color online)  Averaged spin-spin
                  correlations   $\langle   s^x_is^y_j  \rangle$   and
                  $\langle  s^y_is^x_j \rangle$  between spin  $i$ and
                  its     nearest    neighbors     and    next-nearest
                  neighbors.}\label{XYYXBin}
\end{center}
\end{figure}

  There are many different local environments a disk can experience.
We  have therefore studied
  the  average of these  correlations,  defined as
  follows.   We   have  calculated  for  each  site   $i$  its  spin's
  correlation with  its neighbors at  $i+\delta$, where $\delta$  is a
  label running over the N,W,E,S nearest neighbors and NW, SW, SE, and NE
  next-nearest   neighbors (\emph{i.e.}  we  calculate   $\langle   s_i^{\mu}
  s_{i+\delta}^{\nu}\rangle$).   The  site  averages   of  these
  correlation functions were also calculated and the results are shown in 
Figs. \ref{XXYYBin}  and
\ref{XYYXBin} for $N=256$.
In principle there is no need to do an average over disk realizations
 as the site averages are self-averaging quantities.

 There  are  some notable  features  visible  in  these Figures.   The
 correlations  are seen  to grow  as the  packing  fraction increases,
 suggesting that the coupling between spins increases in strength with
 packing  fraction.   Studying  Fig.   \ref{XXYYBin},   the  strongest
 correlations are  seen to be those  with the East and  West spins for
 $\langle s^x_is^x_j \rangle$  and with the North and  South spins for
 $\langle  s^y_is^y_j \rangle$.   It makes sense that  (for example) when a disk is moved
 to the East,  its neighbors to the East and West  should also move in
 that direction.  This will  generate the large  correlations observed
 when $\langle s_i^x s_j^x \rangle$  is measured with the spins to the
 East and West of the central spin.  Studying Fig.  \ref{XYYXBin},  it is
 clear that  for $\langle s^x_is^y_j \rangle$  and $\langle s^y_is^x_j
 \rangle$, the North,  South, East and West correlations  are all zero
 while the others are small but finite. This  confirms the
 presence  of  pseudo-dipolar   interaction  terms  in  the  effective
 Hamiltonian  and is  compatible with  a Hamiltonian
 like that of Eqs. (\ref{hamiltonian}) and (\ref{Ddef}).

\begin{figure}
\begin{center}
	\resizebox{80mm}{!}{\includegraphics{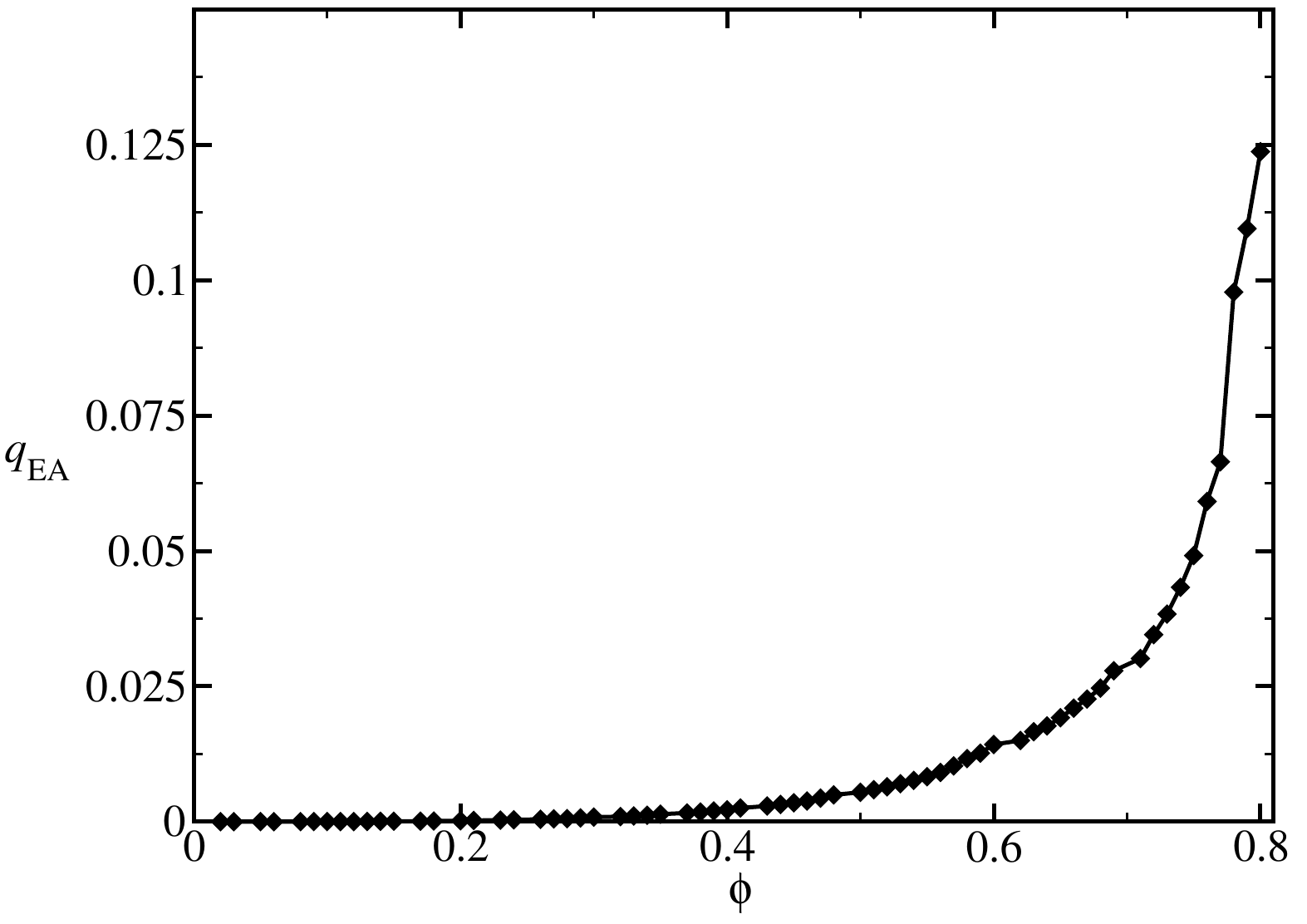}}	
	\caption[Overlap for  the spin system derived  from the binary
          hard disk fluid under  the SOC.]{ Overlap $q$
          measured for a  range of packing fractions in  a system with
          $N=256.$} \label{qALL}
\end{center}
\end{figure}

We have also determined the Edwards-Anderson order (overlap) parameter, defined as
\begin{align}
\displaystyle   q    =   \frac{1}{N}   \sum_{i=1}^{N}\left   ([\langle
  s^x_i\rangle^2]_{av}+[\langle s^y_j\rangle^2]_{av}\right),
\label{qdef}
\end{align}
where  the square brackets  $[\ldots]_{av}$ mean  an average  over the
quenched disorder in  the system (here the species of particle that each cell contains). 
The overlap is a
measure of the  amorphous or glass order in the system.  In Fig.  \ref{qALL},
it can  be  seen that the overlap
increases  as the packing  fraction is  increased towards  its maximum
possible value. It is always non-zero even at small packing fractions.

The overlap measured for a spin glass in a  field
is  finite  at  all  values  of  the  temperature,  growing  larger  as
$T\rightarrow0$.   This   happens  because   the  (random)  fields  bias   the
orientations  of  the spins.

\section{Effective Spin Hamiltonian for hard disks}\label{secHeff}

We   will   try   to    understand   the   correlations   studied   in
Sec. \ref{Correlations} with the  aid of an effective Hamiltonian like
that in Eqs. (\ref{hamiltonian}) and (\ref{Ddef}), but for simplicity
we ignore all couplings except those between nearest neighbors.
This is a
poor approximation at  large packing fractions, but is  better for low
packing fractions. We will also work to lowest non-trivial order for
each quantity studied.

A weak coupling expansion can be made which allows fitting of $A_{ij}$
and $B_{ij}$ from the simulation results.  Unfortunately, as this is a
weak  coupling approximation  (i.e.   it is  valid  when $A_{ij}$  and
$B_{ij}$ are  small) it cannot be  used to accurately  measure them in
the region  of most interest,  $\phi \to \phi_{\text{max}}$,  as there
they become large.

The correlation $<s^{\nu}_is^{\mu}_j>$ is calculated using:
\begin{align}
\displaystyle        \langle        s^{\nu}_is^{\mu}_j\rangle        =
\frac{1}{Z} \int_{-1/2}^{1/2}\int_{-1/2}^{1/2}  s^{\nu}_is^{\mu}_j e^{-\beta
    H_{eff}}                                                    \prod_k
  \,d^{2}s_k,
\end{align}
where the spins components  $s^{\nu}_k$  are integrated over
the $k$th cell, which has unit side length.   The  partition function is
\begin{equation}
Z=\int_{-1/2}^{1/2}\int_{-1/2}^{1/2}e^{-\beta     H_{eff}}    \prod_k
\,d^2s_k.  
\label{partition}
\end{equation}
 The  integrals can  be  performed by  first
Taylor expanding the exponential, and then integrating to
give the correlation  in terms of $A_{ij}$, and  $B_{ij}$ and some simpler
averages. On perfoming the Taylor expansion we find
\begin{align}
\displaystyle \langle  s^{\nu}_is^{\mu}_j \rangle &\approx \frac{1}{Z}
\int_{-1/2}^{1/2}\int_{-1/2}^{1/2}   s^{\nu}_is^{\mu}_j  \left(1-\beta
H_{eff}\right)  \prod_k  \,d^2s_k  \notag  \\  &\approx
\frac{1}{Z}    \int_{-1/2}^{1/2}\int_{-1/2}^{1/2}   s^{\nu}_is^{\mu}_j
\sum_{<lm>} \beta D_{lm}^{\lambda \rho}s_l^{\lambda} s_m^{\rho} \prod_k \,d^2s_k \notag,
\end{align}
and performing the integration yields
\begin{align}\label{BinSpinFit}
 \langle s^{\nu}_is^{\mu}_j \rangle \approx \beta \left[A_{ij} \delta_{\nu
                                                              \mu} - B_{ij}
       \frac{R_{ij}^{\nu}R_{ij}^{\mu}}{|\vec{R}_{ij}|^2}\right]\langle
                     (s^{\nu}_i)^2\rangle\langle(s^{\mu}_j)^2\rangle.
\end{align}
To   the    order   we   are    working   $\langle(s^{\nu}_i)^2\rangle
=\langle(s^{\mu}_i)^2\rangle  \approx 1/12$. 

\begin{figure}
\begin{center}
	\resizebox{80mm}{!}{\includegraphics{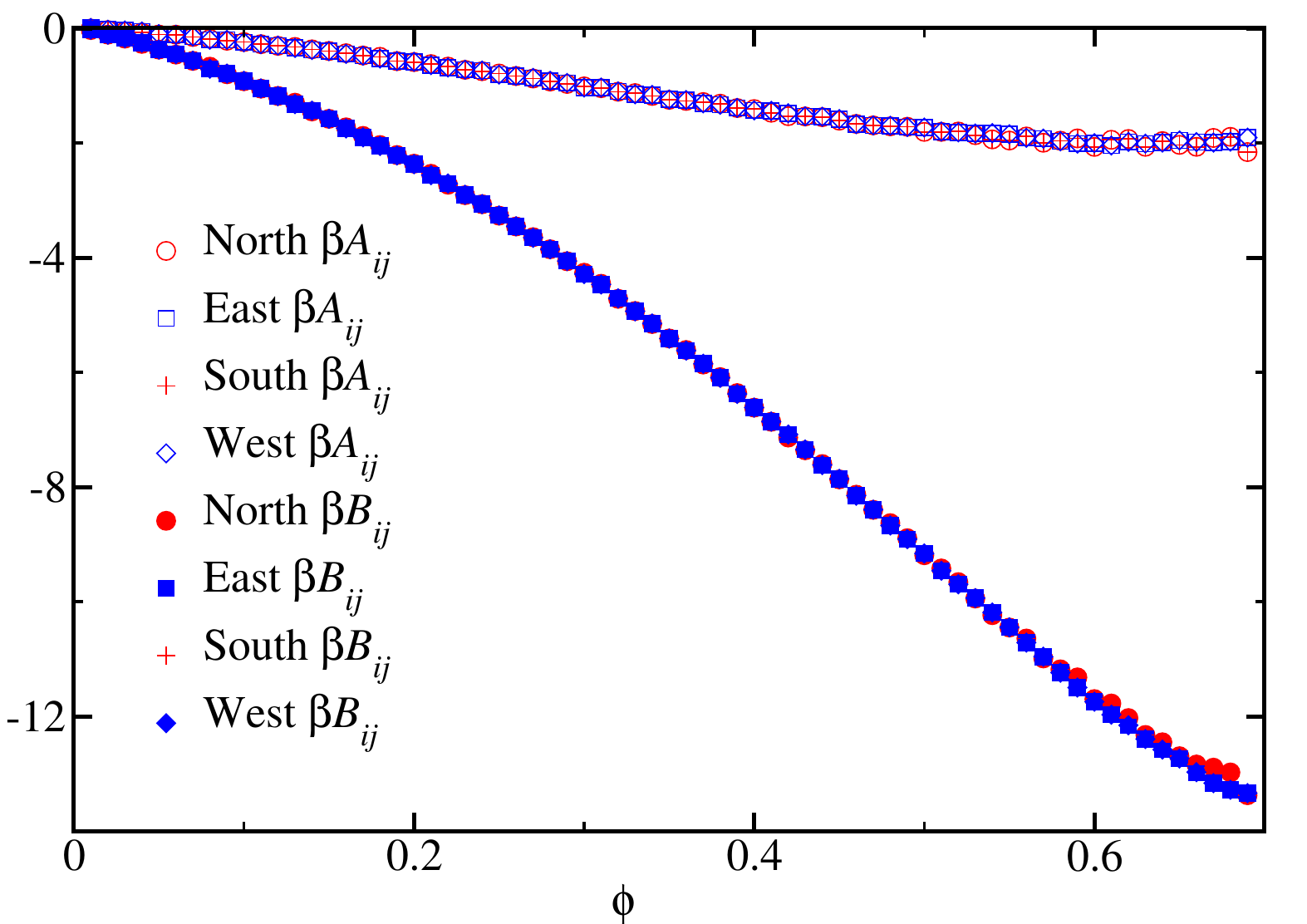}}	
	\caption[Fitting for the  averaged effective coupling constant
          and field strength  for the effective Hamiltonian describing
          the spin derived  from the binary hard disk  fluid under the
          SOC.]{ (Color online) Fitting  of the  averaged $\beta  A_{ij}$  and $\beta
          B_{ij}$  for  a binary  hard  disk  system  under SOC  with
          $N=256$  particles   through  spin-spin  correlations  using
          Eq. (\ref{BinSpinFit}). The four curves for each quantity
          come  from  looking  at  the  North, South,  East  and  West
          directions.  Local  environments  have  been  averaged  over
          leading    to     a    strong    symmetry     between    the
          directions.} \label{BJijAJijB}
\end{center}
\end{figure}

Because we  have measured $\langle s^{\nu}_is^{\mu}_j  \rangle$ we can
use these  measurements to  determine $\beta A_{ij}$  and $\beta
B_{ij}$ for  each bond (nearest-neigbor  pair).  The values  of $\beta
A_{ij}$ (and  $\beta B_{ij}$) have a  distribution, with a  mean and a
standard deviation.  The standard deviation  is important as it is the
randomness of the effective couplings which is encoded in the standard
deviation  which can  be the  cause of  spin glass  behavior if  it is
sufficiently  large  compared  to  the  means of  the  couplings.   In
Fig. \ref{BJijAJijB}  we have plotted  the averages of  $\beta A_{ij}$
and $\beta B_{ij}$ as a function of the packing fraction $\phi$.

Using the same approximation for the effective Hamiltonian we can determine the variance of the random field $\vec{h}_i$ from our results for $q$.
 \begin{align}
\displaystyle        \langle        s^{\mu}_i\rangle        =
\frac{1}{Z}\int_{-1/2}^{1/2}\int_{-1/2}^{1/2}  s^{\mu}_i e^{-\beta
    H_{eff}}                                                    \prod_k
  \,d^2s_k,
\end{align}
and expanding
\begin{align}
\displaystyle \langle  s^{\mu}_i \rangle &\approx \frac{1}{Z}
\int_{-1/2}^{1/2}\int_{-1/2}^{1/2} s_i^{\mu}   \left(1-\beta
H_{eff}\right)  \prod_k  \,d^2s_k  \notag  \\  &\approx
\beta h_i^{\mu} \langle (s_i^{\mu})^2\rangle.
\end{align}
Again one can replace $ \langle (s_i^{\mu})^2\rangle$ by $1/12$.

Thus the variance $h^2$, defined as 
\begin{equation}
(\beta h)^2 =\frac{1}{N}\sum_ i (\beta h_i^{\mu})^2 \approx \frac{144}{N} \sum_i  \langle s_i^{\mu}\rangle^2= 72 q.
\label{hq}
\end{equation}
Eq.   (\ref{hq}) shows  that the  variance  of the  random field  will
increase with packing  fraction just like $q$ does,  at least when $q$
is small, (see  Fig.  \ref{qALL}). The equation will  not hold at high
packing fractions,  where we actually expect $(\beta  h)^2$ to diverge
but $q$  must always remain less than  $\frac{1}{2}$.  (This inequality
arises because  $q$ cannot exceed the  value it would have  if all the
disks were simultaneously at the corners of plaquettes).

\section{The Correlation Lengths}

The quantity  of most interest  is the spin glass  correlation length
as it should be  the glass correlation length.  We  shall
determine it  via the  spin-glass susceptibility.  First  the cumulant
$\chi^{\mu\nu}_{ij}=\langle           s^{\mu}_is^{\nu}_j\rangle-\langle
s^{\mu}_i\rangle\langle s^{\nu}_j\rangle$  is obtained.  This measures
fluctuations  in   the  correlations  between  the   $\mu$  and  $\nu$
components  of  the spins  $i$  and  $j$.  The spin-glass  
wave-vector dependent susceptibility is \cite{SharmaYoung}
\begin{align}
\displaystyle         \chi^{SG}(\vec{k})         =         \frac{1}{N}
\sum_{\mu, \nu}\sum_{i,j}[(\chi^{\mu\nu}_{ij})^2]_{av}\exp(i\vec{k}.\vec{R}_{ij}),
\end{align}
where $\vec{R}_{ij}$ is the vector connecting lattice sites
$i$ and $j$.  From $ \chi^{SG}(\vec{k})$ the spin glass correlation length $\xi^{SG}$ can be
calculated 
using the formula \cite{Young:2004ud}
\begin{align}
\xi^{SG} =
 \frac{1}{2\sin(|\vec{k}_{min}|/2)}\left[\frac{\chi^{SG}(0)}{\chi^{SG}(\vec{k}_{min})}-1\right]^{1/2},
\end{align}
where   $\vec{k}_{min}$   is    the   minimum   non-zero   wave-vector
$\vec{k}_{min}=(2\pi/L,0)$.  

   Additionally,     a    ferromagnetic
correlation length  can be calculated  and compared to the  spin glass
length  to   see  which  kind  of  correlations   are  dominating  the
system. A ferromagnetic wave-vector dependent susceptibility is defined:
\begin{align}
\displaystyle \chi^{F}(\vec{k}) =
 \frac{1}{N} \sum_{\mu, \nu}\sum_{i,j}[(\chi^{\mu\nu}_{ij})]_{av}\exp(i\vec{k}.\vec{R}_{ij}).
\end{align}
This  is similar  to  the  spin glass  susceptibility,  but 
 $\chi^{\mu\nu}_{ij}$ is not squared. From  this, a  ferromagnetic length
scale $\xi^{F}$ can be calculated:
\begin{align}
\xi^{F} = \frac{1}{2\sin(|\vec{k}_{min}|/2)}\left[\frac{\chi^{F}(0)}{\chi^{F}(\vec{k}_{min})}-1\right]^{1/2}.
\end{align}

\begin{figure}
\begin{center}
	\resizebox{80mm}{!}{\includegraphics{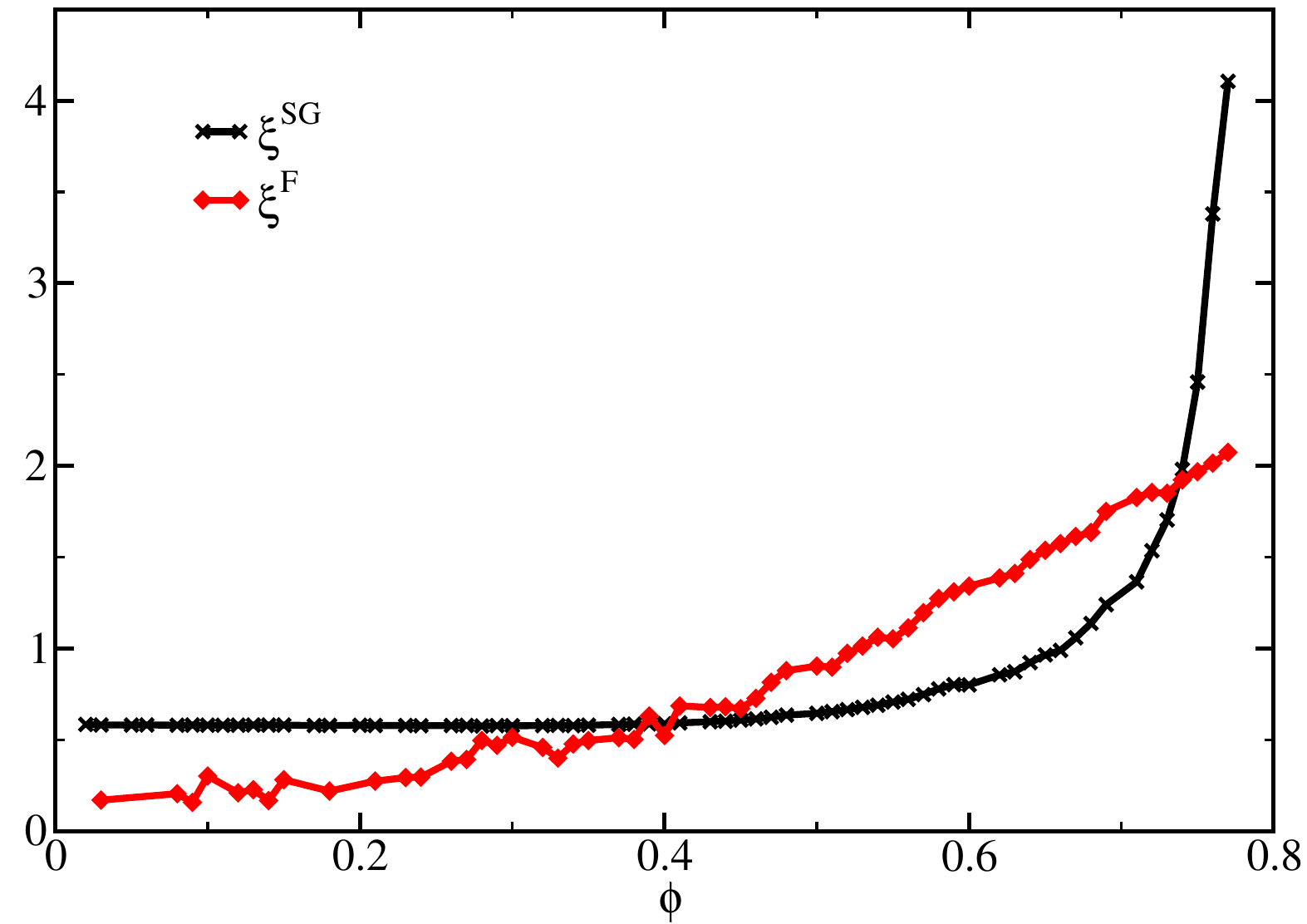}}	
	\caption[ Ferromagnetic and spin glass correlation lengths for
          the  spin system  derived from  the binary  hard  disk fluid
          under the SOC]{ (Color online) Spin glass correlation length
          (in red) and ferromagnetic correlation length (in green) for
          the  spin   system  derived   from  the  binary   hard  disk
          fluid.    Calculated    for    a   system    with    $N=256$
          disks.} \label{CorrLength}
\end{center}\end{figure}

Fig. \ref{CorrLength}  shows these length-scales  plotted together. It
is interesting to  study both cases, since with the  mapping to a spin
system there  is not yet an  \textit {a priori} way  of predicting the
properties of the spin system. There exists another
mapping  of the  structural glass  to a  spin system  by  Stevenson et
al. \cite{Stevenson:2008ko}.  This method is similar to the mapping of
Moore and  Yeo \cite{Yeo},  in that it  requires a replication  of the
structural glass  Hamiltonian but  it results in  a random  bond Ising
model in  random field. The random bond
Ising model contains only couplings $J_{ij}$ of a positive sign, so it
leads to a growing ferromagnetic correlation length rather than a spin
glass correlation  length.  In the  context of the binary  mixture SOC
model, if  the ratio of the  disk sizes $R_{AB}$ gets  close enough to
unity, we would  expect that in this limit, the model  would be in the
universality  class of  the random  field ferromagnet  also.  But this
transition  would be  associated  with the  kind  of crystal  ordering
visible  in  Fig.  \ref{16x16Disks083}  and  seems  irrelevant to  the
physics of glasses.

As  can be  seen  in  Fig. \ref{CorrLength}  at  low packing  fraction
$\xi^{F}$  is  larger  than  $\xi^{SG}$  and  it  grows  with  packing
fraction. However, it does appear to saturate at around 1.5 large disk
radii while $\xi_{SG}$ starts to grow much more rapidly as the packing
fraction  approaches $\phi_{\text{max}}$. This  is good  evidence that
the  important correlations  here are  spin-glass like  and  that when
$R_{AB}=1.0/1.4$ the spin system is not behaving as a ferromagnet in a
random   field.  The   effective  bonds   generated  must   contain  a
sufficiently  large fraction  of  negative bonds  so  that the  system
behaves like a spin glass.

Unfortunately  it  is very  hard  to  measure  the correlation  length
$\xi^{SG}$ well from simulations in  the region of most interest, that
is when $\phi \to  \phi_{\text{max}}$. $\xi^{SG}$ measures the size of
the cooperatively  rearranging regions and  such rearrangements become
very  slow when  the  required rearrangements  involve the  cordinated
motion of many  disks.  This in turn means that  it takes an extremely
long  time for  the system  to equilibrate  and the  susceptibility to
reach its correct level.

\section{Scaling of the effective couplings near $\phi_{\text{max}}$} \label{Maxbehavior}

As $\phi \to \phi_{\text{max}}$ it is just not possible to equilibrate
the system. Furthermore  even if we could measure  the correlations in
this  limit,  we  would  not  be able  to  determine  $h_i^{\mu}$  and
$D_{ij}^{\mu \nu}$ by the procedure of Sec. \ref{secHeff} which relied
on  the validity of  the weak-coupling  approximation, which  fails as
$\phi \to \phi_{\text{max}}$. Our  numerical studies only tell us that
$h_i^{\mu}$  and  $D_{ij}^{\mu   \nu}$  are  increasing  with  packing
fraction.  In  this Section we present  a simple argument  that in the
limit $\phi \to \phi_{\text{max}}$ their dependence  on
packing fraction is as
\begin{equation}
 \beta h_i^{\mu}  \sim 1/(1-\phi/\phi_{\text{max}})
\end{equation}
 and that
\begin{equation}
\beta  D_{ij}^{\mu \nu}  \sim 1/(1-\phi/\phi_{\text{max}})^2.
\end{equation}

 The total phase space of a  finite system of hard disks or spheres is
 fractured into a number of regions (\lq \lq blocked states'') 
which are  mutually inaccessible.  As the density  is pushed up
 there are  fewer and fewer  blocked states \cite{PZ}.   Eventually as
 the packing fraction reaches the maximum for the system there is only
 one blocked state  left. This can be compressed to  a jammed state at
 $\phi_{\text{max}}$.  The pressure  diverges to infinity according to
 Eq.   (\ref{Pdef}).   We  can  use  this observation  to  deduce  how
 $h_i^{\mu}$ and $D_{ij}^{\mu \nu}$  in the effective spin Hamiltonian
 of Eq.  (\ref{hamiltonian}) must vary as $\phi \to \phi_{\text{max}}$
 so as  to recover the exact  expression for the pressure  of the hard
 sphere or hard  disk gas in Eq. (\ref{Pdef}). Our  argument is just a
 variant    of   the    procedure    of   Salsburg    and   Wood
 \cite{Salsburg:1962dd}.

The jammed  state at $\phi_{\text{max}}$  will be isotactic to  a high
degree  of  approximation.  That  is,  each  disk  or sphere  will  be
touching $z  = 2d$ neighbors.  Only a  few (if any) will  be jammed by
virtue of their  centers touching a plaquette wall  and we will assume
this  does  not  occur   for  the  state  at  $\phi_{\text{max}}$.  In
Sec. \ref{Discussion} a variation of  the SOC model based on  Voronoi
cells is outlined  where this will certainly be  true. A finite number
of disks touching the plaquette wall  would not in any case affect the
argument.  Then in  the spin mapping, such a jammed  state should be a
minimum of  the Hamiltonian in Eq.   (\ref{hamiltonian}). Suppose this
minimum occurs at values  of $s_i^{\mu}=S_i^{ \mu}$.  The Hamiltonian
 at its minimum,
\begin{equation}
\beta \mathcal{H}_{min}=-\frac{1}{2}\sum_{i,j}^N (\beta h_i^{\mu})
 F_{ij}^{\mu \nu}
 (\beta h_j^{\nu}),
\end{equation} 
 is just a constant, independent  of $\phi$ with the above scalings of
 $h_i^{\mu}$ and  $D_{ij}^{\mu \nu}$. (Here $F_{ij}^{\mu  \nu}$ is the
 matrix whose  inverse is $ \beta D_{ij}^{\mu  \nu}$). This expression
 for $\beta  \mathcal{H}_{min}$ in  the partition function  defined by
 Eq. (\ref{Zdef}) would not then  give a contribution to the pressure.
 The  pressure is  actually determined  by the  contribution  from the
 \textit{vicinity}  of the  jammed state  at  $\phi_{\text{max}}$.  To
 evaluate    this     contribution    to    $Z$     let    us    write
 $s_i^{\mu}=S_{i}^{\mu}+(1-\phi/\phi_{\text{max}})f_i^{\mu}$.  Because
 we are expanding about a  minimum, the integrals over the $f_i^{\mu}$
 are Gaussian quadratic  forms in the $f_i^{\mu}$ which  do not depend
 on  $(1-\phi/\phi_{\text{max}})$,   with  our  assumed   scalings  of
 $D_{ij}^{\mu  \nu}$.   They  give  a contribution  to  the  partition
 function
\begin{equation}
Z \approx (1-\phi/\phi_{\text{max}})^{dN},
\end{equation}
via the terms which comes from the changes in the integration variable
from $s_i^{\mu}$  to $f_i^{\mu}$.   This yields Eq.   (\ref{Pdef}) for
the pressure. This result is just  a consequence of the scaling assumed for
$h_i^{\mu}$ and $D_{ij}^{\mu \nu}$ with $(1-\phi/\phi_{\text{max}})$.

Note that according  to this argument, both the  mean and the standard
deviation of  the couplings, (which  we generically label  $\beta J_0$
and $\beta  J$ respectively,  without distinguishing the  labels $\mu$
and    $\nu$),   will    scale    in   the    same    way,   viz    as
$1/(1-\phi/\phi_{\text{max}})^2$.     $\beta   h$   will    scale   as
$1/(1-\phi/\phi_{\text{max}})$.

Inserting  these expressions  for $\beta  h$  and $\beta  J$ into  Eq.
(\ref{droplet})  we  recover  Eqs.  (\ref{SGnu}) for  the  correlation
length.   For  hard  disk  systems  and hard  sphere  systems  we  are
therefore  predicting  that  there  is  an actual  divergence  of  the
correlation length as  $\phi \to \phi_{\text{max}}$. The circumstances
where this behavior might  be relevant to  the \textit{unconstrained}
system are discussed below.

\section{Discussion}\label{Discussion}

In a supercooled liquid, a particle  is caged on time scales less than
the  alpha relaxation $\tau_{\alpha}$. On longer time scales it can
diffuse anywhere  in the  system. In the  SOC model, each  particle is
caged forever in  the cell into which it was  first inserted.

We can measure $\tau_{\alpha}$ from the
incoherent scattering function:
\begin{align}
\displaystyle   F(\vec{k},t)  =   \frac{1}{N}  \sum_i   \left  \langle
e^{i\vec{k}.[\vec{r}_i(0)-\vec{r}_i(t)]} \right \rangle.
\end{align}
However, $F(\vec{k},t)$ will  never decay to zero in  the SOC model --
it will fall to  a plateau and remain on the plateau  for all time. In
order  to   see  why  that  happens   consider  the  root-mean-squared
displacement:
\begin{align}
\displaystyle r_{MSD}(t) =
 \left \langle \frac{1}{N}\sum_i \left[\vec{r}_i(t)-\vec{r}_i(0)\right]^2\right\rangle.
\end{align} 
For  the unconstrained  system $r_{MSD}(t)$  first  steadily increases
with time, levels off while the particle  is caged and finally
grows to infinity. For the system
under  the SOC the cell  walls  ensure that
$r_{MSD}(t)$ will  saturate at a value  determined by the  size of the
cell. This  in turn ensures  that $F(\vec{k},t)$ remains  non-zero for
all time.

This does  not mean that  the relaxation  times  of the  SOC system  are
infinite. Consider
\begin{align}
\displaystyle C(t) = \frac{1}{N}\sum_i \left \langle \vec{s}_i(0). \vec{s}_i(t) \right \rangle,
 \end{align}
and note  that $ \left  \langle \vec{s}_i(0) .  \vec{s}_i(t) \right
\rangle  =  \left  \langle  \vec{s}_i \right  \rangle.  \left  \langle
\vec{s}_i \right  \rangle$ as $t$ goes  to infinity, so  in this limit
$C(t)$ approaches  $q$.  The  timescale obtained from  a study  of how
long $C(t)$ takes to reach  $q$ would be similar to $\tau_{\alpha}$ in
the  unconstrained system: the  relaxation time  $\tau_{\alpha}$ comes
about   because  rearrangements  on   the  scale   of  $\xi$   in  the
unconstrained  fluid  are  needed  to  relax  the  cages  holding  the
particles.  In the SOC model, rearrangements on the scale of $\xi$ are
also  required to  allow full  relaxation, so  the two  timescales are
similar.  We leave the details to future studies.

There  is  disorder  present   in  structural  glasses  on  the  alpha
relaxation time scale -- their molecules move so little that the local
environment of  any molecule is effectively  disordered.  However over
periods of many alpha relaxation  times, the disorder is averaged out.
Given this, the SOC model where quenched disorder is built in, may be
appropriate for  studying the behavior  of the fluid on  timescales of
order $\tau_{\alpha}$. Furthermore it  is from data on such timescales
that  one can  obtain estimates  of the  correlation length  $\xi$. We
expect  that at least  when $\xi$  is large  there is  probably little
difference  between  the point-to-set  length  scale  and the  dynamic
length scale \cite{Kob}.

Estimates of the dynamical length scale $\xi$ are obtained as follows.  
The  four-point correlation  function $G_4(\vec{r},t)$   defined as
\cite{Toninelli}:
\begin{eqnarray}\label{4pointcor}
G_4(\vec{r},t)                                                      &  =&
\langle\rho(0,0)\rho(0,t)\rho(\vec{r},0)\rho(\vec{r},t)\rangle \notag \\
&-&\langle\rho(0,0)\rho(0,t)\rangle\langle\rho(\vec{r},0)\rho(\vec{r},t)\rangle,
\end{eqnarray}
 should develop
a plateau when  the liquid starts to become glassy.
The dynamic susceptibility is calculated by integrating
 $G_4(\vec{r},t)$ over volume:
\begin{align}
\chi_4(t) = \frac{1}{V} \int G_4(\vec{r},t) \, d^dr.
\end{align}
When measured in a glassy system, $\chi_4(t)$  is observed to grow with  time, peaking at
times comparable  to $\tau_{\alpha}$ before decaying. As
the temperature is  lowered or the packing fraction  is increased, the
peak moves to  longer and longer times (corresponding  to the increase
in $\tau_{\alpha}$).  The dynamic  susceptibility can be thought of as
a `correlation  volume' which reveals  the scale of regions  which are
dynamically correlated \cite{Berthier:2011tk}, 
providing  evidence of a growing  correlation length  $\xi$  in glassy
systems.

In the  binary disk  SOC system, quenched  disorder is present  in the
form of the random distribution  of disk species over the cells.
The growth of $\chi_{SG}$ and $\xi_{SG}$ reveal
the presence of  growing amorphous order.  Because for  the SOC system the
quenched disorder persists for all time, not just for timescales up to
$\tau_{\alpha}$,  if $\chi_4(t)$  were measured  in the  SOC  model it
would grow and then saturate at $\chi_{SG}$.

It  is our  belief  that SOC  models  of hard  disks  and spheres  can
therefore  describe the increase  of $\xi$  with packing  fraction, at
least as regards the value of the exponent $\nu$. We do not expect the
value of $\phi_{\text{max}}$ to  necessarily coincide with the packing
fraction of the  divergence in the unconstrained system  -- after all,
$\phi_{\text{max}}$ would be of  slightly different value if the cells
had not  had a square shape  or even were  of random shape.  A  way of
constructing \lq \lq random'' cells  would be to equilibrate
the binary disk or sphere system and then use as the cells the Voronoi
cells of a single configuration as the cells. Because of this
built-in randomness,  this same procedure  could be used to  model the
striking  glassy features  of monodisperse  spheres. (For  the Voronoi
cell version of the SOC model, the argument in Sec.  \ref{Maxbehavior}
is clearly exact \cite{Salsburg:1962dd}.   On the other hand, for such
cells it would be impossible  to carry out the analytical calculations
in  Sec. \ref{spinH}).  

The  divergence of  $\xi$ as  $\phi \to  \phi_{\text{max}}$ is  likely  to be
accompanied by a  divergence of the relaxation time  of the SOC model.
Note  that  such   behavior  is  not  that  expected   of  a  G  point
\cite{BerthierWitten}. At a G point $\xi$ and $\tau$ both diverge, but
the pressure remains finite. At $\phi_{\text{max}}$ the pressure is infinite,
as it  is also  a jammed  state.

 Our value for $\phi_{\text{max}}$ is  quite close to the estimates of
 the value of the random close-packing fraction: $\phi_{\text{rcp}} \approx 0.84$
 \cite{Teitel}.   We think that  this similarity  is not  an accident.
 Both    the   packing    fractions,    $\phi_{\text{max}}$   and    $
 \phi_{\text{rcp}}$,  are  obtained from  situations  where the  phase
 space of the hard disks has  been curtailed so that the system cannot
 stray far from its initial state.  $\phi_{\text{max}}$ will depend on
 the choices made for the shape  of the cells.  It will also depend on
 how the large  and small particles are assigned to  the cells. In our
 work  this  has been  done  randomly but  one  could  build into  the
 distribution if  desired the local correlations  of the unconstrained
 system.   It is  also known  that the  random close  packing fraction
 $\phi_{\text{rcp}}$ is not well-defined: it has a small dependence on
 the protocol used to determine it \cite{Teitel}.

When studying the unconstrained hard  sphere or hard disk system, some
protocol has  to be adopted to  see glassy behavior, such  as a finite
compression  rate, and  this  will  result in  the  pressure going  to
infinity  at some  packing  fraction  less than  that  of the  densest
crystalline  state.   In  true  equilibrium, the  pressure  of  course
remains  finite unless the  system is  at the  maximum density  of the
crystalline state.   We believe that  the glass state  is well-defined
provided that  the alpha relaxation time $\tau_{\alpha}$  is such that
$1/\tau_{\alpha}$ is  greater than the  rate for phase  separation and
crystallization in the case of  binary mixtures, or the time scale for
crystal nucleation  and growth  generally.  A finite  compression rate
should not modify the  quasi-equilibrium approach to glasses (like that
in   this   paper)   provided   that   it   is   small   compared   to
$1/\tau_{\alpha}$.  Since the alpha relaxation time $\tau_{\alpha}$ is
expected  to grow with  $\xi$ as  $\ln \tau_{\alpha}  \sim \xi^{\psi}$
\cite{dynamics}, then for  a fixed compression rate one  can only hope
to  obtain the growth  of $\xi$  up to  a compression  rate determined
value.   But within  these various  constraints we  believe  the glass
problem  is well  defined and  that  SOC models  are a  useful way  of
studying some aspects of it.

We would  expect the  SOC model of  hard disks  or spheres to  be most
useful at  densities above $\phi_d$,  the density at  which timescales
increase rapidly.  This density can be quite  well-understood with the
aid of  mode-coupling theory.  For hard spheres  $\phi_d \approx 0.58$
and for  hard disks  $\phi_d \approx 0.78$  \cite{PZ}. In the  case of
hard disks  in the  SOC model, timescales  were seen to  increase very
rapidly at a rather similar density. This is because at such densities
the timescales are long because they involve collective rearrangements
of the disks on a length  scale $\xi$ and collisions with the walls of
the  plaquette  are becoming insignificant. Alas,  this  very  rapid
increase   makes numerical investigations   at densities  above
$\phi_d$ very challenging.

\acknowledgements{  We   should  like   to  thank  Les   Woodcock  for
  introducing us to SOC models and Mike Godfrey for many discussions of them.
 One of us (CJF) acknowledges financial
  support while in Manchester of an EPSRC doctoral studentship.}


\begin{thebibliography}{99}

\bibitem{Berthier05} L. Berthier, G. Biroli, J.-P. Bouchaud, L. Cipelleti, D. El Masri, D. L. H\"{o}te, F. Ladieu and M. Pierno, Science, {\bf 310}, 1797 (2005).

\bibitem{Cavagna} A. Cavagna, T. S. Grigera and P. Verrocchio, Phys. Rev. Lett. {\bf 98},187801 (2007).

\bibitem{Biroli} G. Biroli, J.-P. Bouchaud, A. Cavagna, T. S. Grigera and 
P. Verrocchio, Nature Phys. {\bf 4}, 771 (2008).

\bibitem{Kob}W. Kob, S. Rold\'{a}n-Vargas, and L. Berthier, Nature Phys. {\bf 8}, 164  (2012).

\bibitem{Berthier:2011tk} L. Berthier, Physics {\bf 4}, 42 (2011).

\bibitem{Kurchan} J. Kurchan and D. Levine, J. Phys. A.: Math. Theor. {\bf 44}, 035001 (2011).

\bibitem{KTW} T. R. Kirkpatrick and P. G. Wolynes, Phys.\ Rev.\ A {\bf
35}, 3072 (1987); T. R.  Kirkpatrick and D. Thirumalai, Phys.\ Rev.\ B
{\bf 36}, 5388 (1987); T.  R.  Kirkpatrick and P.  G.  Wolynes, Phys.\
Rev.\ B {\bf 36}, 8552 (1987).

\bibitem{LW} V. Lubchenko and P. G. Wolynes,  Annu. Rev. Phys. Chem. {\bf 58}, 235 (2007).

\bibitem{BB} G. Biroli and J.-P. Bouchaud, arXiv:0912.2542.

\bibitem{MP}M. M\'{e}zard and G. Parisi, Phys. Rev. Lett.
  {\bf  82}, 747 (1999).
\bibitem{PZ} G. Parisi and F. Zamponi,  Rev. Mod. Phys. {\bf 82},  789 (2010).

\bibitem{Kauzmann} W. Kauzmann, Chem. Rev. \textbf{ 43}, 219 (1948).

\bibitem{YM12b} J. Yeo and M. A. Moore, Phys. Rev. E {\bf 88}, 052501 (2012).

\bibitem{MD} M. A. Moore and B. Drossel, Phys.\ Rev.\ Lett.\ {\bf 89}, 217202 (2002).

\bibitem{YM} J. Yeo and M. A. Moore,  Phys. Rev. B  {\bf 85}, 100405(R) (2012).

\bibitem{TM} M. Tarzia and M. A. Moore, Phys. Rev. E {\bf 75}, 031502 (2007).

\bibitem{Yeo}M.~A.~Moore and J.~Yeo, Phys. Rev. Lett. {\bf 96}, 095701 (2006).

\bibitem{Hoover:1966cf} W. G. Hoover and B. J. Alder, J. Chem. Phys. {\bf 45}, 2361 (1962).

\bibitem{Hoover:1968ux} W. G. Hoover and F. H. Ree, J. Chem. Phys. {\bf 49}, 3609 (1968).

\bibitem{Nelson} D. R. Nelson and B. I. Halperin, Phys. Rev. B {\bf 19}, 2457 (1979).

\bibitem{AT} J. R. L. de Almeida and D. J. Thouless, J. Phys. A {\bf 11}, 983 (1978).

\bibitem{Moore12} M. A. Moore, Phys. Rev. E {\bf 86}, 031114 (2012).

\bibitem{AT6} M. A. Moore and A. J. Bray, Phys.\ Rev.\ B {\bf 83}, 224408 (2011).

\bibitem{LS} B. D. Lubachevsky and F. H. Stillinger, J. Stat. Phys. {\bf 60}, 561 (1990).

\bibitem{McM} W. L. McMillan, Phys. Rev. B {\bf 29}, 4026 (1984).

\bibitem{BM} A. J. Bray and M. A. Moore, Lecture Notes in Physics, {\bf 275}, 121 (1986).

\bibitem{FH} D. S. Fisher and D. A. Huse, Phys. Rev. Lett. {\bf 56}, 1601 (1986);
 Phys. Rev. B {\bf 38}, 386 (1988);\textit{ibid}. {\bf 38}, 373 (1988).

\bibitem{Carter} A. K. Hartmann, A. J. Bray, A. C. Carter, M. A. Moore,
 and A. J. Young, Phys. Rev. B {\bf 66}, 224401 (2002).

\bibitem{Boettcher} S. Boettcher, Phys. Rev. Lett. {\bf 95}, 197205 (2005).

\bibitem{Salsburg:1962dd} Z. W. Salsburg and W. W. Wood, J. Chem. Phys. {\bf 37}, 798 (1962).

\bibitem{Fisher:1965br} M. E. Fisher, J. Chem. Phys. {\bf 42}, 3852 (1965).

\bibitem{Henley:1993ul} C. N. Likos and C. L. Henley, Phil. Mag. B {\bf 68}, 85 (1993).

\bibitem{Teitel} D. Vagberg, D. Valdez-Balderas, M. A. Moore, P.  Olsson, 
and S.  Teitel, Phys. Rev. E {\bf 83}, 030303 (2011).

\bibitem{Perera:1999hd} D. N. Perera and P. Harrowell, Phys. Rev. E  {\bf 59}, 5721 (1999).

\bibitem{Hentschel:2007ig} H. G. E. Hentschel, V. Ilyin, N. Makedonska, I. Procaccia,
 and N. Schupper, Phys. Rev. E {\bf 75}, 050404 (2007).

\bibitem{Hentschel:2008} H. G. E. Hentschel, V. Ilyin and I. Procaccia, Phys. Rev. Lett. {\bf 101}, 265701 (2008).

\bibitem{Dyre} S. Toxvaerd, U. R.  Pedersen, T. B.  Schroder, J. C. Dyre, J. Chem. Phys. {\bf 130}, 224501 (2009).

\bibitem{Godfreystatic} M. J. Godfrey and M. A. Moore, in preparation.

\bibitem{flocks} W. Bialek, A. Cavagna, I. Giardina, T. Mora, E. Silvestri, M. Viale, and
 A. M. Walczak,  PNAS {\bf 27}, 4786 (2012).

\bibitem{BrayMoore82} A. J. Bray and M. A. Moore, J. Phys. C: Solid State Physics, {\bf 15}, 3897 (1982).

\bibitem{SharmaYoung} A. Sharma and A. P. Young, Phys. Rev. E {\bf 81}, 061115 (2010).

\bibitem{Moore06} M. A. Moore, Phys. Rev. Lett. {\bf 96}, 137202 (2006).

\bibitem{Allen:1987wd} M. P. Allen and D. Tildesley, \textit{Computer Simulations of Liquids}, (Oxford Scientific Publications, Oxford, 1987).

\bibitem{Lubachevsky:829677} B. D. Lubachevsky, J. Comput. Phys. {\bf 94}, 255 (2005).

\bibitem{Torquato10} S. Torquato and F. H. Stillinger, Rev. Mod. Phys.
 {\bf 82},2633 (2010).

\bibitem{Young:2004ud} A. P. Young and H. G. Katzgraber, Phys. Rev. Lett. {\bf 93}, 207203 (2004).

\bibitem{Stevenson:2008ko} J. D. Stevenson, A. M. Walczak, R. W. Hall, and
 P. G. Wolynes, J. Chem. Phys. {\bf 129}, 194505 (2008).

\bibitem{Toninelli} C. Toninelli, M.  Wyart, L.  Berthier, G. Biroli, J.-P.  Bouchaud, Phys. Rev. E {\bf 71}, 0141505 (2005).

\bibitem{BerthierWitten} L. Berthier and T. A. Witten, Phys. Rev. {\bf E}, 021502 (2009).

\bibitem{dynamics} M. Barnett-Jones, M. J. Godfrey, T. Grundy and M. A. Moore, 
cond-mat arXiv: 1211.1915.

\end{thebibliography}
\end{document}